\documentclass[amsmath,amssymb,aps,prd,nofootinbib,letterpaper,superscriptaddress,longbibliography,notitlepage]{revtex4-1}

\usepackage[usenames,dvipsnames]{color}
\usepackage{graphicx,epstopdf}
\usepackage{tensor}
\usepackage{subcaption}
\usepackage[bookmarks]{hyperref}
\usepackage{amsmath}
\usepackage{array}
\newcolumntype{M}[1]{>{\centering\arraybackslash}m{#1}}
\usepackage{mathtools}
\usepackage{zref-savepos}
\usepackage[normalem]{ulem}

\newcommand{\doublerule}[1][.4pt]{%
  \noindent
  \makebox[0pt][l]{\rule[1ex]{\linewidth}{#1}}%
  \rule[.3ex]{\linewidth}{#1}}
  
\begin{document}

\title{Extended I-Love relations for slowly rotating neutron stars}

\author{J\'er\'emie Gagnon-Bischoff}
\email{jgagn129@uottawa.ca}
\affiliation{Department of Physics, University of Ottawa, 75 Laurier Ave E, Ottawa, Ontario, K1N 6N5, Canada}
\author{Stephen R. Green}
\email{stephen.green@aei.mpg.de}
\affiliation{Max Planck Institute for Gravitational Physics (Albert Einstein Institute), Am M\"uhlenberg 1, 14476 Potsdam, Germany}
\author{Philippe Landry}
\email{landryp@uchicago.edu}
\affiliation{Enrico Fermi Institute and Kavli Institute for Cosmological Physics, The University of Chicago, 5640 South Ellis Avenue, Chicago, Illinois, 60637, USA}
\author{N\'estor Ortiz}
\email{nortiz@perimeterinstitute.ca}
\affiliation{Perimeter Institute for Theoretical Physics, 31 Caroline Street North, Waterloo, Ontario, N2L 2Y5, Canada}

\date{\today}

\begin{abstract}
  Observations of gravitational waves from inspiralling neutron star
  binaries---such as GW170817---can be used to constrain the nuclear
  equation of state by placing bounds on stellar tidal
  deformability. For slowly rotating neutron stars, the response to a
  weak quadrupolar tidal field is characterized by four
  internal-structure-dependent constants called ``Love
  numbers". The \emph{tidal} Love numbers $k_2^\text{el}$ and
  $k_2^\text{mag}$ measure the tides raised by the gravitoelectric and
  gravitomagnetic components of the applied field, and the
  \emph{rotational-tidal} Love numbers $\mathfrak{f}^\text{o}$ and
  $\mathfrak{k}^\text{o}$ measure those raised by couplings between
  the applied field and the neutron star spin. In this work we compute
  these four Love numbers for perfect fluid neutron stars with
  realistic equations of state. We discover (nearly) equation-of-state
  \emph{independent} relations between the rotational-tidal Love
  numbers and the moment of inertia, thereby extending the scope of
  I-Love-Q universality. We find that similar relations hold among the
  tidal and rotational-tidal Love numbers. These relations extend the
  applications of I-Love universality in gravitational-wave
  astronomy. As our findings differ from those reported in the
  literature, we derive general formulas for the rotational-tidal Love
  numbers in post-Newtonian theory and confirm numerically that they
  agree with our general-relativistic computations in the weak-field
  limit.
\end{abstract}

\maketitle

\section{Introduction}\label{sec:intro}

The Advanced LIGO and Virgo gravitational-wave observatories have
recently detected the inspiral and merger of two low-mass compact objects consistent with 
neutron stars (NSs)~\cite{GW170817}. With three operating detectors and the
simultaneous observation of a gamma-ray burst, it was possible to
localize the event in the sky and perform follow-up observations of
electromagnetic counterparts. Together, these exciting observations
have provided a host of insights into neutron star physics, gamma-ray
bursts, kilonovae, and even cosmology.

The inspiral stage of the binary NS merger, seen only by the
gravitational wave detectors, can provide insight into the internal
structure and composition of the NSs through their tidal
deformability. As the stars inspiral, they exert tidal forces on each
other, resulting in deformed stars. This affects the orbital dynamics
of the binary, and slightly accelerates the
coalescence~\cite{Lai,Kochanek,Flanagan}. Indeed, through
nonobservation of an unambiguous tidal phase shift, Ref.~\cite{GW170817}
placed bounds on the tidal deformability of the stars and thereby
constrained competing nuclear physics models of the NS interior.

Cold NS matter is described as a barotropic perfect
fluid with a particular equation of state (EoS) determined by the
underlying nuclear physics model. In recent years, much effort has
been devoted to describing the tidal deformability of compact objects
in general relativity, including NSs, as a function of their
EoS~\cite{Hinderer,Favata,Damour_Nagar,Binnington,Surficial,Irrotational}.
The main result of this program is that for weak and slowly varying
tides the dependence on the EoS is captured by a set of constants
called \emph{Love numbers}. The Love numbers relate the induced field
of the tidally deformed body to the applied tidal field.

The analysis of Ref.~\cite{GW170817} constrained only the
leading-order tides raised by the presence of the binary companion,
but as detector sensitivity is improved it will become important to
have waveform predictions that also take into account tides raised by
the orbital motion of the companion, as well as interactions between
tidal fields and NS spin. The main purpose of this work is to compute
and study the Love numbers associated with this rotational-tidal
response for slowly rotating NSs with realistic EoSs. Our main result
is to show that existing universal relations between NS observables
can be extended to include these Love numbers. These relations could
be helpful for modelling and measuring spin corrections to tidal
effects in the waveform, although the computation of such corrections
is beyond the scope of this work.

In general relativity, the applied tidal field is described by two
sets of symmetric-tracefree (STF) tensors \cite{Zhang}, the
gravitoelectric and gravitomagnetic multipole moments
$\{\mathcal{E}_L\}_{\ell \geq 2}$ and
$\{\mathcal{B}_L\}_{\ell \geq 2}$, where $L = a_1 a_2 ... a_{\ell}$ is
a spatial multi-index. These tidal moments are assumed to be sourced
by an external mass and momentum distribution, such as a binary
companion. At the linear level, neglecting the spin of the body, an
$\ell$-pole gravitoelectric tidal field will induce as a tidal
response a mass moment
$k_{\ell}^{\text{el}} R^{2\ell+1} \mathcal{E}_L/G$, while an
$\ell$-pole gravitomagnetic tidal field will induce a current moment
$k_{\ell}^{\text{mag}} M R^{2\ell} \mathcal{B}_L/c^2$ (modulo
normalizations), where $M$ and $R$ are the mass and radius of the body.
We refer to $k_{\ell}^{\text{el}}$ and
$k_{\ell}^{\text{mag}}$ as the gravitoelectric and gravitomagnetic
tidal Love numbers. The
gravitoelectric tidal Love numbers reduce to the Love numbers of
Newtonian theory in the weak-field limit, while the gravitomagnetic
tidal Love numbers arise only in general relativity. Together, they
provide a complete description of the deformation of a
\emph{nonrotating} body subject to a weak, slowly varying tidal
field. The tidal Love numbers have been calculated for a variety of
configurations, including polytropes
\cite{Hinderer,Damour_Nagar,Binnington,Favata,Irrotational}, NSs with
realistic EoSs
\cite{Read,Hindereretal,DamourNagarVillain,Read2,Maselli} and quark
stars \cite{Postnikov,Lau}, and they have been shown to vanish
identically for black holes \cite{Binnington}. Several works have
studied their impact on binary NS waveforms, showing a slight speed-up
in the
merger~\cite{Read,Hindereretal,Baiotti,Baiotti2,Vines,Pannarale,Lackey,DamourNagarVillain,Read2,Maselli,Lackey2}.

Neutron stars in binaries generically have \emph{nonzero spin}
$S^a$. This couples nonlinearly to the applied tidal field to generate
additional corrections to the gravitational-wave phase. Observations
indicate, however, that the dimensionless spin
$\chi^a\equiv cS^a/GM^2$ is
small---$|\chi^a|\lesssim0.05$~\cite{Hannam}---so the effect is likely
to be suppressed relative to the leading-order pure tidal
deformation~\cite{Pani}. Assuming low spin and rigid rotation, the
leading-order rotational-tidal couplings are proportional to
$\chi^a\mathcal{E}_L$ and $\chi^a\mathcal{B}_L$. These couplings
combine the dipole spin vector with an $\ell$-pole tidal moment,
generating \emph{bilinear moments} of multipole orders
$\ell-1 \leq \ell' \leq \ell+1$. Just as a tidal Love number measures
the amplitude of the gravitational field induced by a given tidal
moment, a \emph{rotational-tidal Love number} measures the field
induced by a bilinear moment.

In this paper we restrict to \emph{quadrupolar} applied tides. While
spin couplings with the tidal quadrupole moments generate $\ell=1,2,3$
bilinear moments, the rotational-tidal response is in fact fully
characterized by two octupole rotational-tidal Love numbers,
$\mathfrak{f}^{\text{o}}$ (gravitoelectric)\footnote{Note that we
redefine this scaled Love number relative to Landry and
Poisson~\cite{External_metric,Internal_metric}:
$\mathfrak{f}^{\text{o}}[\text{here}] \equiv
\mathfrak{f}^{\text{o}}[\text{LP}] + \frac{5}{3}k_2^{\text{el}}$
(see Sec.~\ref{SubSec:f_o}). Our new definition coincides with that
of Ref.~\cite{Pani} up to an overall scale.} and
$\mathfrak{k}^{\text{o}}$ (gravitomagnetic). Indeed, there are no
$\ell = 1$ Love numbers, as a dipole deformation represents an overall
acceleration of the body, which can be eliminated by switching to its
center-of-mass frame. Moreover, the $\ell = 2$ Love numbers identified
in Refs.~\cite{External_metric,Dynamical_response,Internal_metric} can
be eliminated by transforming to a suitably rotated frame, as we show
in Appendix~\ref{SubSec:QuadLNs}. Thus, the response of a rotating
body to quadrupolar tides is characterized by the four Love numbers
$\{k_2^{\text{el}}, k_2^{\text{mag}},
\mathfrak{f}^{\text{o}},\mathfrak{k}^{\text{o}}\}$. These vanish for
black holes~\cite{External_metric}, while for material bodies they
depend on the EoS. Throughout this paper we will also use versions of
the Love numbers scaled by powers of stellar compactness $GM/c^2 R$. These will
be denoted by uppercase letters
$ K_2^\text{el}~\equiv~(c^2R/2GM)^5 k_2^\text{el}$,
$\mathfrak{F}^{\text{o}} \equiv -(c^2R/2GM)^5
\mathfrak{f}^{\text{o}}$,
$K_2^\text{mag} \equiv (c^2R/2GM)^4 k_2^\text{mag}$, and
$\mathfrak{K}^{\text{o}} \equiv -(c^2R/2GM)^5 \mathfrak{k}^{\text{o}}
$. The scaled Love numbers are the quantities that enter into the
universality relations~\cite{ILQ}, whereas the genuine Love numbers
remain finite and nonzero in the zero-compactness limit,
$GM/c^2R \rightarrow 0$~\cite{External_metric}.

Section~\ref{Sec:TidScales} is devoted to justifying our restriction
to quadrupolar applied tides. We show how the various tidal fields and
Love numbers appear in the metric and we identify some of their
physical effects. Supposing the tidal fields are sourced by a binary
companion, we estimate the size of each term, and we argue that higher
multipole terms make a smaller contribution to the metric. We note,
however, that the higher-$\ell$ terms could still contribute
significantly to the waveform itself, and this should be investigated
in future work.

A complete analysis of the deformation of a slowly rotating body
subject to a quadrupolar tidal field was carried out by Landry and
Poisson
\cite{slow_BH,External_metric,Dynamical_response,Internal_metric}, and
separately by Pani, Gualtieri, Maselli and Ferrari
\cite{Pani_formalism,Pani}, who also investigated the effect of an
octupolar tidal field and worked to second order in spin. The two
frameworks differ primarily in their assumptions about the fluid
state: Pani, Gualtieri and Ferrari hold the fluid completely
static~\cite{Pani}, while Landry and Poisson allow it to develop tidal
currents~\cite{Irrotational} in accordance with the circulation
theorem of relativistic hydrodynamics~\cite{Synge}. Because these
fluid motions are vorticity-free in a nonrotating star, the latter
state has been termed \emph{irrotational}. The static
state is incompatible with the Einstein equation except in
axisymmetry~\cite{Dynamical_response}. In this work we follow the
framework of Landry and Poisson, which we review in
Sec.~\ref{Sec:LoveNumbers}.

In Sec.~\ref{Sec:results}, we compute the four Love numbers for
polytropes and an incompressible fluid, and we confirm that they agree
with results of independent post-Newtonian calculations in the
zero-compactness limit. We find that our results do not match those of
Ref.~\cite{Pani}, even in the regime where we expect fluid state
differences to be insignificant. In particular, we observe no evidence
of a sign change of $\mathfrak{F}^{\text{o}}$ (their
$\delta \tilde{\lambda}_M^{32}$) at high
compactness~\cite{Pani}. The $n=1$ polytrope
results presented in Ref.~\cite{Pani} are also incompatible with our
independent post-Newtonian calculation.

The octupole rotational-tidal Love numbers have been calculated for
polytropes by Landry~\cite{Internal_metric}, and for realistic-EoS NSs
in the static fluid state by Pani, Gualtieri and Ferrari~\cite{Pani}.
In Sec.~\ref{Sec:RealEoS}, we perform the first-ever computation of
$\mathfrak{f}^{\text{o}}$ and $\mathfrak{k}^{\text{o}}$ for realistic NSs
in the irrotational state. The EoSs we adopt are piecewise polytropic
fits to a subset of the candidate EoSs considered in
Ref.~\cite{Read_etal09}. Our models are chosen to be compatible with
the maximum observed NS mass of approximately
$2M_\odot$~\cite{Demorest,Antoniadis}, and they also respect the
causal bound on the sound speed. We find that these realistic EoSs
have Love numbers that lie between those of the $n=0.5$ and $n=1$
polytropes, and we observe a qualitative difference between $npe\mu$
and exotic-matter EoSs.

One remarkable property of NSs is the existence of approximately
EoS-independent relations between three macroscopic quantities: the
moment of inertia $I$, the scaled Love number $K_2^{\text{el}}$, and
the rotational quadrupole moment $Q$. Although each quantity depends
on the EoS in a separate way, Yagi and Yunes~\cite{ILQ} discovered
empirical functional relationships between them that are almost
completely insensitive to the EoS. The origin of this \emph{I-Love-Q
  universality} is not entirely understood, although it has been
linked to the emergence of isodensity contour self-similarity in
compact stars~\cite{kYlSgPnYtA14}.  I-Love-Q relations have important
applications in gravitational-wave astronomy.  They imply that a
precise measurement of one element of the triad is sufficient to
determine the other two with percent-level accuracy. This can break
the degeneracy between spin-spin interaction and rotational quadrupole
contributions to the phasing of NS binary waveforms; an independent
measurement of $K_2^{\text{el}}$ could permit the extraction of the
individual NS spins~\cite{kYnY13}. Universal relations also have
applications in parameter estimation and strong-field tests of general
relativity~\cite{ILQreview}. Recent work has revealed that similar
EoS-independent relations exist between the scaled gravitomagnetic and
gravitoelectric tidal Love numbers \cite{ILQmulti}, between the moment
of inertia and the scaled gravitomagnetic tidal Love number
$K_2^{\text{mag}}$ \cite{Delsate}, and between the scaled tidal Love
numbers of two NSs in a binary system~\cite{ILQbinary,ILQtidparams},
among other combinations (see Ref.~\cite{ILQreview} for a review).

In Sec.~\ref{Sec:Universality} we extend I-Love universality to
include the scaled rotational-tidal Love numbers. We show that these
\emph{extended I-Love relations} hold to within $2.5\%$ accuracy. This
contrasts with Ref.~\cite{Pani}, which found deviations from
universality of up to 200\%. Extended I-Love universality also
suggests the existence of universal relations among the scaled Love
numbers.  We confirm explicitly that these \emph{Love-Love} relations
hold between each pair of the four scaled Love numbers we study. Our
findings provide further evidence that NSs can be characterized in an
approximate way by a single number (beyond the mass and the spin) that
determines their internal-structure dependent properties.

We also include several appendices. Appendix~\ref{SubSec:PN_calc}
describes our calculations of Love numbers in the post-Newtonian
approximation. While general post-Newtonian formulas for the tidal
Love numbers are known in the literature~\cite{Irrotational}, only
partial results exist for the rotational-tidal Love numbers; here, we
derive expressions for $\mathfrak{f}^{\text{o}}$ and
$\mathfrak{k}^{\text{o}}$ that are valid for any barotropic EoS. In
Appendix~\ref{SubSec:Incompress}, we adapt the recipe presented in
Sec.~\ref{Sec:LoveNumbers} to treat an incompressible fluid, which is
a limiting case in terms of stiffness of the EoS. In
Appendix~\ref{SubSec:QuadLNs} we show that the aforementioned
quadrupole rotational-tidal Love numbers are spurious. And in
Appendix~\ref{SubSec:PGF} we derive the mapping between our Love
numbers and those of Ref.~\cite{Pani}.

Throughout this manuscript, lower case Latin indices $a, b, c, \ldots$
denote spatial components, and are raised and lowered with the
Euclidean metric $\delta_{ab}$. Upper case Latin indices
$A, B, C, \ldots$ denote angles $\theta^A \equiv (\theta,\phi)$, and
are raised and lowered with the $\mathcal{S}^2$ metric
$\mathcal{S}_{AB} \equiv \text{diag}(1,\sin^2{\theta})$. Greek indices
represent spacetime components. With the exception of
Secs.~\ref{sec:intro}, \ref{Sec:TidScales} and
Appendix~\ref{SubSec:PN_calc}, we work in geometrized units
$G=c=1$.

\section{Tidal scales} \label{Sec:TidScales}

The rotational-tidal couplings we study in this paper are generated by
the quadrupole moments $\mathcal{E}_{ab}$, $\mathcal{B}_{ab}$ of the
tidal field. We work to first order in $\chi^a$, $\mathcal{E}_{ab}$
and $\mathcal{B}_{ab}$, and we also treat the second order bilinear
terms $\chi^a \mathcal{E}_{bc}$ and $\chi^a \mathcal{B}_{bc}$. We
claim that the bilinear octupole deformations associated with
$\chi^a \mathcal{E}_{bc}$ and $\chi^a \mathcal{B}_{bc}$ represent
important subleading corrections to the leading-order tides raised
directly by the gravitoelectric quadrupole moment
$\mathcal{E}_{ab}$. To justify this, in this section we describe the
various scales of the problem, and we determine the relative sizes of
the deformations induced by the tidal field.

We consider a body of mass $M$, radius $R$ and dimensionless spin
angular momentum $\chi \equiv |\chi^a| \ll 1$ in a vacuum region of spacetime
pervaded by the tidal influence of distant mass and momentum
distributions. We imagine that these distributions are sourced by a
binary companion of mass $M_{\text{tid}}$ at a separation of
$b \gg R$. (Alternately, $M_{\text{tid}}$ and $b$ can be interpreted
as generic mass and distance scales for the tidal source.) To leading
order, the companion generates quadrupolar tidal fields
$\mathcal{E}_{ab}$, $\mathcal{B}_{ab}$. These appear in the spacetime
metric with the scalings \cite{Vlasov}
\begin{equation} \label{scaling}
\frac{r^2 \mathcal{E}_{ab}}{c^2} \sim \frac{G M_{\text{tid}}}{c^2 b} \left( \frac{r}{b} \right)^2 , \qquad \frac{r^2 \mathcal{B}_{ab}}{c^3} \sim \frac{G M_{\text{tid}}}{c^2 b} \left(\frac{v}{c}\right) \left( \frac{r}{b} \right)^2 ,
\end{equation}
where $r$ measures distance from the body's center of mass, $r \ll b$
in the neighborhood of the body and\linebreak
${v\sim\sqrt{G(M+M_{\text{tid}})/b}}$ is a velocity scale for the
companion's orbital motion. We neglect the higher multipoles of the
tidal field, such as the tidal octupole moments $\mathcal{E}_{abc}$,
$\mathcal{B}_{abc}$, because they are suppressed by $(r/b)$ relative
to the tidal quadrupole moments.

The couplings between $\chi^a$ and the tidal quadrupole moments
produce terms of the form $\chi^a \mathcal{E}_{bc}$,
$\chi^a \mathcal{B}_{bc}$ in the metric; they are suppressed by a
factor of $\chi$ relative to those in Eq.~\eqref{scaling}. When
$\chi \gg r/b$, these bilinear terms dominate over the octupole tidal
terms. In this case, the further assumption that
$r \sim G(M+M_{\text{tid}})/c^2$ in the neighborhood of the body
implies that the dimensionless spin satisfies
$v^2/c^2 \ll \chi \ll 1$. This condition is naturally fulfilled when
the binary separation is large and the component masses are broadly comparable. (If $\chi < v^2/c^2$, the bilinear terms are negligible and the tidal octupole terms are the important subleading corrections.)

By examining the form of the metric far from the tidally perturbed,
slowly rotating body, we can determine the relative sizes of the
deformations associated with each of the tidal and bilinear
moments. The generic external metric ansatz was constructed by Landry
and Poisson~\cite{External_metric,Internal_metric}. In an expansion in
powers of $GM/c^2r$, the time-time and time-angle components of the
metric are
\begin{subequations} \label{schem}
\begin{align} \label{gtta}
g_{tt} =& -1 + \frac{2GM}{c^2r} - \left[ 1 + \ldots + 2 k_2^{\text{el}} \left( \frac{R}{r} \right)^5 \left( 1 + \ldots \right) \right] \frac{\mathcal{E}_{ab} x^a x^b}{c^2} + \frac{2GM}{c^2}\left(1+\ldots\right)\frac{\chi^b \mathcal{B}_{ab}x^a}{c^3} \nonumber \\  &- \frac{2GM}{c^2 r^2} \left[ \frac{GM}{c^2 r} + \ldots + 2 \mathfrak{k}^{\text{o}} \left(\frac{R}{r}\right)^5 \left( 1+ \ldots \right)\right] \frac{\chi_{\langle a}\mathcal{B}_{bc \rangle} x^a x^b x^c}{c^3} , \\ \nonumber
g_{tA} =& \, \frac{2G^2M^2}{c^4r^3} \epsilon_{abc} x^b \chi^c x^a_A + \frac{2}{3} \left[ 1 + \ldots - 6 \left( \frac{GM}{c^2 r} \right) k_2^{\text{mag}} \left( \frac{R}{r} \right)^4 \left( 1 + \ldots \right) \right] \frac{\epsilon_{acd} x^c \mathcal{B}^d_{\;\;b} x^b x^a_A}{c^3} \\ &- \frac{2G^2M^2}{c^4 r} \left( 1 + \ldots \right) \frac{\epsilon_{abc} x^b \mathcal{E}^c_{\;\;d} \chi^d x^a_A}{c^2} - \frac{10 G M}{3 c^2 r^2} \left[ \frac{G^2 M^2}{c^4 r} + \ldots + \frac{6}{5} \mathfrak{f}^{\text{o}} \left(\frac{R}{r}\right)^5 \left( 1 + \ldots \right) \right] \frac{\epsilon_{ac}^{\;\;\;\, d} x^c \mathcal{E}_{\langle db} \chi_{e \rangle} x^b x^e x^a_A}{c^2} . \label{gttb}
\end{align}
\end{subequations}
in the Regge-Wheeler gauge and Boyer-Lindquist $(t,r,\theta,\phi)$
coordinates. Here, ellipses denote relativistic corrections of order
$GM/c^2r$ and higher, $x^a$ are Cartesian mass-centred coordinates,
$x^a_A\equiv \partial x^a/\partial\theta^A$ are their angular
derivatives, and $\epsilon_{abc}$ is the antisymmetric permutation
symbol.

In $g_{tt}$ the set of terms proportional to $\mathcal{E}_{ab}x^a x^b$
describes a quadrupole deformation of the spacetime. The leading term
in square brackets represents the applied gravitoelectric field, while
the decaying term proportional to the Love number $k_2^{\text{el}}$ is
the tidal response. This deformation is a Newtonian effect at leading
order, as indicated by the factor of $c^{-2}$. The $tt$ component also
contains a dipole deformation resulting from the coupling of the
body's spin to the gravitomagnetic part of the tidal field; this is an
overall acceleration of the body due to the Mathisson-Papapetrou spin
force \cite{Mathisson,Papapetrou,Corinaldesi}, which enters as an
order 1.5 post-Newtonian (1.5PN) correction by virtue of the factor of
$c^{-5}$ \cite{External_metric}. The last set of terms in $g_{tt}$,
proportional to
$\chi_{\langle a}\mathcal{B}_{bc \rangle} x^a x^b x^c$, describes the
octupole deformation of the spacetime due to another coupling of
$\chi^a$ and $\mathcal{B}_{ab}$. The decaying piece of this
solution---the tidal response---involves the rotational-tidal Love
number $\mathfrak{k}^{\text{o}}$. This octupole deformation is a 1.5PN
effect which is suppressed relative to the deformation associated with
$k_2^{\text{el}}$ by a factor of $\chi(v/c) GM/c^2 r$.

Turning to the $tA$ component of the metric, the terms proportional to
$\epsilon_{acd} x^c \mathcal{B}^d_{\;\;b} x^b$ represent a
gravitomagnetic quadrupole deformation of the spacetime. The
gravitomagnetic field is itself a 1PN phenomenon (as indicated by the $c^{-3}$ scaling in this time-space component), but the tidal
response measured by $k_2^{\text{mag}}$ is suppressed by an additional
factor of $GM/c^2r$ in the metric. The deformation associated with
$k_2^{\text{mag}}$ is therefore a 2PN effect, smaller than the
gravitoelectric quadrupole deformation by a factor of $(v/c)GM/c^2
r$. The set of terms proportional to
$\epsilon_{ac}^{\;\;\;\, d} x^c \mathcal{E}_{\langle db} \chi_{e
  \rangle} x^b x^e$ in $g_{tA}$ corresponds to an octupole deformation
generated by the coupling of the spin to the external gravitoelectric
field; it enters at 1.5PN and is suppressed by $\chi GM/c^2 r$
relative to the deformation associated with $k_2^{\text{el}}$. The
amplitude of the decaying, tidal-response piece in square brackets is
determined by $\mathfrak{f}^{\text{o}}$. The $tA$ component also
contains a term proportional to $\epsilon_{abc} x^b \chi^c$ describing
the body's rotation, as well as a 2.5PN dipole deformation proportional to $\epsilon_{abc}
x^{b} \mathcal{E}^{c}_{\phantom{c}d} \chi^{d}$ due to
another spin force.

The structure of the metric allows us to remark on the expected
scaling of the bilinear quadrupole deformations that would have been
induced by couplings between $\chi^a$ and $\mathcal{E}_{abc}$,
$\mathcal{B}_{abc}$, had we included the tidal octupole moments in our
description. Because such bilinear terms are generated via octupole
(rather than quadrupole) couplings, they will be suppressed relative
to their counterparts in Eq.~\eqref{schem} by a factor of $r/b$,
although they are of the same post-Newtonian order (1.5PN).

On the basis of these scalings, we conclude that the most important
corrections to the leading-order tides measured by $k_2^{\text{el}}$
come from the deformations associated with $k_2^{\text{mag}}$ and
$\mathfrak{f}^{\text{o}}$. They are smaller than the quadrupole
induced by $\mathcal{E}_{ab}$ by factors of $(v/c)(GM/c^2r)$ and
$\chi (GM/c^2r)$, respectively. Although $\chi \gg v^2/c^2$, the
comparison of $\chi$ and $v/c$ is ambiguous, and depends on the
parameters of the binary. Very early during binary inspiral,
\begin{equation}
  \frac{v}{c} \sim \frac{\left[G(M+M_{\text{tid}}) f_{\text{GW}}\right]^{1/3}}{c} \approx 0.06 \left( \frac{M+M_{\text{tid}}}{2M_{\odot}} \right)^{1/3} \left( \frac{f_{\text{GW}}}{25 \text{~Hz}} \right)^{1/3} ,
\end{equation}
where $f_{\text{GW}}$ is twice the orbital frequency; for a
lightweight binary of total mass~$\sim 2M_\odot$, this is comparable
to the maximum known spin for
a NS in a binary that merges within the Hubble time,
$\chi \approx 0.05$~\cite{Hannam}. Hence, the deformations produced by
rotational-tidal couplings can be just as large as the tides due to
gravitomagnetism.

We remark that the preceding discussion does not mean that
$k_2^{\text{mag}}$ and $\mathfrak{f}^{\text{o}}$ necessarily make the
largest contributions (after $k_2^{\text{el}}$) to the
gravitational-wave tidal \emph{phase}. Determining the precise
corrections the rotational-tidal Love numbers make to the tidal
phasing of the binary NS waveform is beyond the scope of this work.

\section{Framework}\label{Sec:LoveNumbers}
In this section we describe the approach of Landry and
Poisson~\cite{External_metric,Dynamical_response,Internal_metric} for
treating the deformation of a slowly and rigidly rotating NS caused by a weak,
slowly varying quadrupolar tidal field. The idea is to
solve the Einstein-fluid equations for a NS subject to
asymptotic conditions corresponding to the applied field rather than
the standard asymptotically flat conditions. One can then read off the
induced field, whose amplitude determines the Love number, from the solution.

We consider a four-dimensional spacetime described by a metric
tensor $g_{\alpha\beta}$, and we treat the NS matter as a perfect
fluid with energy-momentum tensor
\begin{equation}\label{eq:Tab}
T_{\alpha\beta} = (\mu + p)u_\alpha u_\beta + p \, g_{\alpha \beta}.
\end{equation}
Here, $\mu$ and $p$ are the total fluid energy density and pressure,
and $u^\alpha$ is the four-velocity of the fluid elements. The total
energy density $\mu$ is the sum of the rest mass density $\rho$ and
the internal thermodynamic energy $\epsilon$. We assume the fluid to
be barotropic, with EoS $p=p(\rho)$. The remaining fluid state
variables follow from the EoS and the first law of thermodynamics for
barotropic fluids,
\begin{equation}\label{eq:first_law}
d(\epsilon/\rho) = -p \, d(1/\rho).
\end{equation}
The matter and metric satisfy the Einstein equation and, for a
one-parameter EoS, all of the hydrodynamic equations follow from
energy-momentum conservation.

For weak tides and slow rotation, the spacetime and matter fields
describing the NS and its neighborhood differ by a small amount from
those of an isolated, nonrotating NS. This allows us to work in
perturbation theory about a static, spherically symmetric
background star. We describe the background solution in the following
subsection.

In Sec.~\ref{SubSec:Perturbed_metric} we write down the form of the
tidally perturbed metric. This ansatz is constructed by adding terms
to the background metric proportional to the moments of the applied
tidal field and the spin of the star to interpolate between the star
and the tidal environment, with radial dependence to be determined by
the Einstein equation. We keep terms proportional to
$\mathcal{E}_{ab}$, $\mathcal{B}_{ab}$, $\chi^a$, as well as the
bilinear quantities $\chi^a \mathcal{E}_{bc}$ and
$\chi^a \mathcal{B}_{bc}$.

To solve the Einstein equation it is convenient to split the problem
into two parts, the interior and exterior regions of the star. We
present the exterior solution (known analytically) in
subsection~\ref{SubSec:Exterior_solution}. The exterior solution
matches to the applied field far away, but also contains subleading
parts with undetermined coefficients, the four scaled Love
numbers
$\{ K_2^\text{el}, K_2^\text{mag}, \mathfrak{F}^{\text{o}},
\mathfrak{K}^{\text{o}} \}$. These quantities are determined by
matching to the interior solution, which is required to be regular at
the origin. We describe the matter part of the interior solution in
Sec.~\ref{SubSec:Perturbed_matter}, and the solution for the
metric---as well as the procedure for obtaining the Love numbers---in
Sec.~\ref{SubSec:Interior_solution}. The interior solution will
typically be determined numerically, and it depends on the chosen
fluid EoS.

We assume throughout that the dynamical time scale
$\sqrt{b^3/(M+M_\text{tid})}$ of the tidal field is much longer than
the characteristic time scale $\sqrt{R^3/M}$ of the internal stellar
dynamics. This is a physically reasonable assumption for the inspiral
stage of a binary NS system when the orbital separation is many times
larger than the stellar radius. Accordingly, we treat the applied tides as
stationary.

\subsection{Background solution}\label{SubSec:Unperturbed_spacetime}
The background solution is taken to be a static, spherically symmetric
star with line element
\begin{equation}\label{eq:unperturbed_metric}
\overline{ds}^2 = - e^{2 \psi(r)} dt^2 + f(r)^{-1} dr^2 + r^2d\mathcal{S}^2,
\end{equation}
where $f(r) \equiv 1 - 2m(r)/r$ and $d \mathcal{S}^2 \equiv \mathcal{S}_{AB} d\theta^A d\theta^B$. The functions $\psi(r)$ and $m(r)$ are
determined from the matter by the Einstein equation, which reduces to
two ordinary differential equations (ODEs),
\begin{eqnarray}
  \frac{dm}{dr} &=& 4 \pi r^2 \bar{\mu}, \label{eq:simplified_EFE1}\\
  \frac{d\psi}{dr} &=& \frac{m + 4 \pi r^3 \bar{p}}{r^2 f}.\label{eq:simplified_EFE2}
\end{eqnarray}
Here, and in the rest of Sec.~\ref{Sec:LoveNumbers}, we use overbars to denote background quantities. Conservation of energy-momentum, along with the above equations, yields
the Tolman-Oppenheimer-Volkoff (TOV) equation,
\begin{equation}\label{eq:TOV}
\frac{d\bar{p}}{dr} = - \frac{(\bar{\mu} +\bar{p})(m + 4 \pi r^3 \bar{p} )}{r^2 f}.
\end{equation}
The equations of structure Eqs.~\eqref{eq:simplified_EFE1}--\eqref{eq:TOV} are
completed by the EoS. Outside the star, where $T_{\alpha\beta}=0$, the solution is
Schwarzschild, with $f(r) = e^{2 \psi(r)} = 1 - 2M/r$. Inside the star ($0\le r \le R$), the equations of structure are solved subject to the regularity condition $m(0) = 0$ at the center, and the matching conditions $\bar{p}(R) = 0$ and $m(R)=M$ at the surface.

\subsection{Perturbed metric ansatz}\label{SubSec:Perturbed_metric}

Following the detailed analysis of
Refs.~\cite{External_metric,Dynamical_response,Internal_metric}, the
metric describing the tidally deformed, slowly rotating star is
constructed by adding deformation terms to the unperturbed metric~\eqref{eq:unperturbed_metric}. To describe the pure tidal
response, we add terms proportional to the applied quadrupolar tidal
field $\mathcal{E}_{ab}$ and $\mathcal{B}_{ab}$. To describe the slow
rotation we add a term proportional to the angular velocity
$\Omega^a \equiv \chi^aM^2/I$, where $I$ is the moment of inertia. And
to study the rotational-tidal response we add terms proportional to
the bilinear quantities $\Omega^a\mathcal{E}_{bc}$ and
$\Omega^a\mathcal{B}_{bc}$. These mixed terms are decomposed with
respect to parity and multipole order (dipole, quadrupole, octupole)
into the bilinear moments defined in Table~\ref{tb:bili_mom}. All the
terms are multiplied by functions of $r$ to be determined later using
the field equations. The radial functions are designed to encapsulate
the star's tidal response, and ensure that at large $r$ the metric
asymptotes to that of the tidal environment \cite{Vlasov}.

The precise form of the perturbed metric is constructed such that it
transforms suitably under parity and rotations. To do this, all
moments---tidal, rotational, rotational-tidal---are repackaged into a
set of \emph{potentials} by taking duals and contracting with the unit
radial vector $n^a\equiv x^a/r$ and its angular derivatives
$n^a_A \equiv \partial n^a/\partial\theta^A$ (see
Table~\ref{tb:bili_pot}). These are inserted according to their
transformation properties in the various components of the metric. In
Regge-Wheeler gauge, the perturbed metric takes the form
\begin{subequations}
  \label{eq:interior_metric}
  \begin{alignat}{6}
    g_{tt} =& -e^{2\psi(r)} + e_{tt}^\text{q}(r)\mathcal{E}^\text{q} + k_{tt}^\text{d}(r)\mathcal{K}^\text{d} + k_{tt}^\text{o}(r)\mathcal{K}^\text{o}, \\
    g_{tr} =& ~ \hat{e}_{tr}^\text{q}(r)\hat{\mathcal{E}}^\text{q} + k_{tr}^\text{d}(t,r)\mathcal{K}^\text{d}+k_{tr}^\text{o}(t,r)\mathcal{K}^\text{o} , \\
    g_{rr} =& ~ f^{-1} + e_{rr}^\text{q}(r)\mathcal{E}^\text{q}+ k_{rr}^\text{d}(r)\mathcal{K}^\text{d} + k_{rr}^\text{o}(r)\mathcal{K}^\text{o}, \\
    g_{tA} =& ~[1-\omega(r)]r^2 \Omega_A^\text{d} +b_t^\text{q}(r) \mathcal{B}_A^\text{q}+\hat{b}_{t}^\text{q}(t,r) \hat{\mathcal{B}}_A^\text{q} + f_t^\text{d}(r)\mathcal{F}_A^\text{d}+ f_t^\text{o}(r)\mathcal{F}_A^\text{o}, \\
    g_{rA} =& ~ \hat{b}_{r}^\text{q}(r) \hat{\mathcal{B}}_A^\text{q}, \\
    g_{AB} =& ~ r^2 \mathcal{S}_{AB} + e^\text{q}(r) \mathcal{S}_{AB} \mathcal{E}^\text{q} + k^\text{o}(r)\mathcal{S}_{AB}\mathcal{K}^\text{o}.
  \end{alignat}
\end{subequations}
This form of the metric ensures that the perturbed Einstein equation
will automatically decompose according to the potentials. Close
examination of the metric shows that the star's spin appears only in
$g_{tA}$, the pure gravitoelectric tide appears in the diagonal components, and
the pure gravitomagnetic tide appears only in
$g_{tA}$~\cite{Binnington}.

Most of the coefficients of the potentials in
Eq.~\eqref{eq:interior_metric} are functions of $r$ alone as a
consequence of our assumption of stationary tides. However, some of
them acquire a time dependence through gravitomagnetic induction
inside the rotating star, even when $\mathcal{B}_{ab}$ is
stationary~\cite{Dynamical_response}. Without the internal dynamics,
the nonaxisymmetric part of the tidal response would violate the
Einstein equation. Following
Refs.~\cite{Dynamical_response,Internal_metric}, we assume that the time dependence of the
metric perturbations associated with $k_{tr}^{\text{d}}$, $\hat{b}^{\text{q}}_{t}$, and
$k^{\text{o}}_{tr}$ can be at most linear in $\Omega t$. The field equations then show that
\begin{subequations} \label{eq:time_dependence}
  \begin{eqnarray} 
    k^{\text{d}}_{tr}(t,r) &=& t k^{\text{d}}_{tr1}(r),\\
    \hat{b}^{\text{q}}_{t}(t,r) &=& t  \hat{b}^{\text{q}}_{t1}(r),\\
    k^{\text{o}}_{tr}(t,r) &=& t k^{\text{o}}_{tr1}(r).
  \end{eqnarray}
\end{subequations}
While time dependence of some kind is physically required, the linear growth is an artifact of the perturbative expansion. Indeed, the time dependence has been shown to be bounded in a physical setting \cite{Internal_metric}. The functions~\eqref{eq:time_dependence} vanish outside the star, so the exterior solution
remains stationary.

\begin{table}
  \begin{ruledtabular}
    \centering
    \begin{tabular}{ M{0.25\textwidth} M{0.25\textwidth} M{0.25\textwidth} M{0.25\textwidth} }
      Moment & Definition & Parity & Multipole order $\ell$ \\ 
      \hline   \noalign{\medskip}
      $\mathcal{F}_a$ & $\mathcal{E}_{ab}\Omega^b$ & Odd & 1 \\ 
      $\hat{\mathcal{E}}_{ab}$ & $2\Omega^c\epsilon_{cd(a}\mathcal{E}^d_{\phantom bb)}$ & Even & 2 \\ 
      $\mathcal{F}_{abc}$ & $\mathcal{E}_{\langle ab}\Omega_{c\rangle}$ & Odd & 3 \\ 
      \noalign{\medskip}
      $\mathcal{K}_a$ & $\mathcal{B}_{ab}\Omega^b$ & Even & 1 \\ 
      $\hat{\mathcal{B}}_{ab}$ & $2\Omega^c\epsilon_{cd(a}\mathcal{B}^d_{\phantom bb)}$ & Odd & 2 \\ 
      $\mathcal{K}_{abc}$ & $\mathcal{B}_{\langle ab}\Omega_{c\rangle}$ & Even & 3 \\ 
    \end{tabular}
  \end{ruledtabular}
  \caption{Bilinear moments resulting from couplings of the dipole
    angular velocity vector $\Omega^a$ to the quadrupolar tidal
    field. Parentheses indicate symmetrization and angular brackets
    indicate the STF operation (symmetrization and removal of all
    traces).}
  \label{tb:bili_mom}
\end{table}
\begin{table}
\begin{ruledtabular}
  \centering
  \begin{tabular}{ M{0.5\textwidth} M{0.5\textwidth} }
    Potential & Definition \\ 
    \hline 	\noalign{\medskip}
    $\Omega^\text{d}_A$ & $\epsilon_{abc}n^b\Omega^c n^a_A$ \\ \medskip
    $\mathcal{E}^\text{q}$ & $\mathcal{E}_{ab}n^a n^b$ \\
    $\mathcal{B}^\text{q}_A$ & $\epsilon_{abc}n^b \mathcal{B}_{\phantom dd}^c n^d n^a_A$ \\ \medskip
    $\mathcal{F}_A^\text{d}$ & $\epsilon_{abc}n^b\mathcal{F}^c n^a_A$ \\  
    $\hat{\mathcal{E}}^\text{q}$ & $\hat{\mathcal{E}}_{ab}n^a n^b$ \\
    $\hat{\mathcal{E}}^\text{q}_A$ & $(\delta_a^{\;\,b} - n_a n^b)\hat{\mathcal{E}}_{bc}n^c n^a_A$ \\
    $\mathcal{F}_A^\text{o}$ & $\epsilon_{abc}n^b\mathcal{F}^c_{\phantom dde} n^d n^e n^a_A$ \\  \medskip
    $\hat{\mathcal{B}}^\text{q}_A$ & $\epsilon_{abc}n^b \hat{\mathcal{B}}_{\phantom dd}^c n^d n^a_A$ \\ 
    $\mathcal{K}^\text{d}$ & $\mathcal{K}_{a}n^a$ \\
    $\mathcal{K}^\text{d}_A$ & $ (\delta_a^{\;\,b} - n_a n^b) \mathcal{K}_{b}n^a_A$ \\
    $\mathcal{K}^\text{o}$ & $\mathcal{K}_{abc}n^a n^b n^c$ \\
    $\mathcal{K}^\text{o}_A$ & $ (\delta_a^{\;\,d} - n_a n^d) \mathcal{K}_{dbc} n^b n^c n^a_A$ \\
  \end{tabular}
\end{ruledtabular}
\caption{Potentials appearing in the metric and fluid ansatzes of
  Eqs.~\eqref{eq:interior_metric} and
  \eqref{eq:perturbed_fluid}. The construction
  of the tidal potentials is described in Ref.~\cite{Vlasov} and the
  bilinear potentials in Refs.~\cite{slow_BH,External_metric}.}
\label{tb:bili_pot}
\end{table}

\subsection{Exterior solution}\label{SubSec:Exterior_solution}

Outside the star, the energy-momentum tensor vanishes. Using the metric
ansatz~\eqref{eq:interior_metric}, and discarding terms of second or
higher order in spin or tides, the vacuum Einstein equation decouples
according to the potentials into ODEs. These can then be integrated to
obtain \emph{analytic} expressions for the radial functions
\cite{External_metric}.

The function $\omega$ satisfies the ODE
\begin{equation}
  r \frac{d^2\omega}{dr^2} + 4 \frac{d\omega}{dr} = 0,
\end{equation}
and we choose $\omega = 1 - 2I/r^3$ as the solution so that, in the
absence of tides, the exterior reduces to the linearized Kerr
spacetime. The solution contains a free parameter, the moment of
inertia $I$, which must be determined by matching to the interior
solution.

The equations for the radial functions associated with the tidal
potentials $\mathcal{E}^{\text{q}}$ and $\mathcal{B}^{\text{q}}_A$
reduce to two second-order homogeneous ODEs. One of them determines the
function $e^{\text{q}}_{tt}$, which is algebraically related to the
other gravitoelectric-sector functions
$\{e_{rr}^{\text{q}},e^{\text{q}}\}$. The other governs the sole
gravitomagnetic-sector function $b^{\text{q}}_t$.

For each ODE, there exist two independent solutions: one decaying in
powers of $r$, the other growing. The amplitude of the growing
solution is set so that the spacetime outside the star matches onto
the tidal environment at large $r$. The amplitude of the
decaying solution---the tidal Love number---is set by matching to the regular
interior solution at $r=R$. The exterior solutions for
$e^{\text{q}}_{tt}$ and $b^{\text{q}}_{t}$, involving undetermined scaled
Love numbers $K_2^{\text{el}}$ and $K_2^{\text{mag}}$, are listed in
Table~\ref{tb:external_metric}.

The radial functions associated with the bilinear potentials satisfy
second-order \emph{inhomogeneous} ODEs sourced by the functions
$e^{\text{q}}_{tt}$, $b^{\text{q}}_{t}$ and $\omega$, which generate
particular solutions in addition to the growing and decaying
ones. Nevertheless, there are still two free parameters in each
exterior solution, and they are set by the boundary conditions at the
stellar surface and at large $r$. In this case, however, the
coefficient of a decaying solution is not necessarily a Love number:
some of the constants are pure gauge \cite{External_metric} (see
Appendix~\ref{SubSec:QuadLNs} for an example). By carefully
identifying and eliminating the gauge constants, one is left with two
ODEs for the octupole radial functions: the first determines
$f^{\text{o}}_t$, and the second determines $k^{\text{o}}_{tt}$, which
is algebraically related to
$\{k_{rr}^{\text{o}},k^{\text{o}}\}$. Analytic expressions for these
functions---involving the undetermined scaled rotational-tidal Love numbers
$\mathfrak{F}^{\text{o}}$ and $\mathfrak{K}^{\text{o}}$---are given in
Table~\ref{tb:external_metric}. The functions
$\{ k^{\text{d}}_{tr}, \hat{b}^{\text{q}}_{t}, k^{\text{o}}_{tr} \}$
vanish by virtue of the vacuum Einstein equation.

The result of this discussion is that only a subset
$\{e^{\text{q}}_{tt}, b^{\text{q}}_t, k^{\text{o}}_{tt},
f^{\text{o}}_t\}$ of the external radial functions are needed to
compute the Love numbers. These functions appear solely in the $tt$
and $tA$ components of the metric. For the complete exterior solution
to the problem, we refer the reader to Ref.~\cite{External_metric}.

\begin{table}
  \footnotesize
  \doublerule
    \begin{flalign*}
      &e_{tt}^\text{q} = -4M^2x^2 \left\{ \left(1-\frac{1}{x}\right)^2 + \frac{2}{x^5} \left[-30x^3(x-1)^2\ln{\left(1-\frac{1}{x}\right)} - \frac{5}{2} x(2x-1)\left(6x^2-6x-1\right) \right] K_2^\text{el}\right\} 
      \\ \medskip
      &b_t^\text{q}=\frac{16M^3x^3}{3}\bigg\{\left(1-\frac{1}{x}\right)-\frac{3}{x^5} \left[20x^4(x-1)\ln{\left(1-\frac{1}{x}\right)} + \frac{5}{3}x\left(12x^3-6x^2-2x-1\right)\right] K_2^\text{mag}\bigg\} 
      \\ \medskip
      &k_{tt}^\text{o} = -4Ix^2 \bigg\{ \frac{1}{x^7} \left[-10x^4(x-1)\left(280x^3-420x^2+140x+3\right)\ln{\left(1-\frac{1}{x}\right)} - 2800x^7+5600x^6-\frac{9100}{3}x^5 + \frac{610}{3}x^4 +\frac{115}{3} x^3 + 5x^2 -\frac{5}{6}x \right.\\ 
      \medskip
      & \qquad \left. -\frac{5}{6} \right] K_2^\text{mag} + \frac{2}{x^6} \left[-420x^4(2x-1)(x-1)^2\ln{\left(1-\frac{1}{x}\right)} -7x^2\left(120x^4-240x^3+130x^2-10x-1\right) \right] \mathfrak{K}^\text{o} +\frac{1}{2x^2} - \frac{1}{2x^3} \bigg\} 
      \\ \medskip
      &f_t^\text{o} = 8IMx^3 \bigg\{\frac{5}{4x^7}\left[4x^3\left(420x^5-700x^4+280x^3+5x-2\right)\ln{\left(1-\frac{1}{x}\right)} + \frac{2}{3}x^2\left(2520x^5-2940x^4+420x^3+70x^2+44x+3\right)\right] K_2^\text{el} 
      \\
      & \qquad +\frac{2}{x^6}\left[210x^5(3x-2)(x-1)\ln{\left(1-\frac{1}{x}\right)}+\frac{7}{2}x^2\left(180x^4-210x^3+30x^2+5x+1\right)\right]\mathfrak{F}^\text{o} -\frac{5}{12x^3}+\frac{1}{6x^4} \bigg\} 
    \end{flalign*}
  \doublerule
  \captionsetup{justification=raggedright,singlelinecheck=off}
  \caption{Select radial functions appearing in the $tt$ and $tA$
    components of the exterior metric, expressed in terms of
    $x = r/(2M)$. All functions within square brackets behave as
    $1+ {\cal O}(1/x)$ when $x\gg1$. We remark that our expression for
    $f^{\text{o}}_t$ differs from that of
    Refs.~\cite{External_metric,Internal_metric} because of our
    redefinition of $\mathfrak{F}^{\text{o}}$ (see
    Sec.~\ref{SubSec:f_o}).}
  \label{tb:external_metric}
\end{table}
\normalsize

\subsection{Perturbed fluid}\label{SubSec:Perturbed_matter}

The interior solution is governed by the hydrodynamic equations. In this
subsection, we use these equations and the Einstein equation to cast the fluid variables in
terms of the radial functions from the metric ansatz.

We begin by decomposing the perturbed fluid variables $\mu$, $p$,
$u_r$ and $u_A$ in terms of the tidal and bilinear potentials of
Table~\ref{tb:bili_pot}, as was done for the metric ansatz. (The time
component of the fluid four-velocity is automatically fixed by
properly normalizing $u^\alpha$.) The decomposition is presented in
detail in Refs.~\cite{Dynamical_response, Internal_metric}.

We next impose energy-momentum conservation. For an unperturbed
barotrope satisfying the first law of
thermodynamics~\eqref{eq:first_law}, this reduces to the conservation
statement $\mathcal{L}_u \omega_{\alpha\beta} = 0$, where
$\mathcal{L}_u$ is a Lie derivative along~$u^{\alpha}$. In other
words, the vorticity tensor
$\omega_{\alpha\beta} \equiv \partial_{\alpha}(h u_{\beta}) -
\partial_{\beta}(h u_{\alpha})$ is conserved along the fluid
wordlines. Here, $h\equiv(\mu+p)/\rho$ is the
specific enthalpy. Taking a variation, and assuming that the EoS is
unchanged by the perturbation, it follows that
\begin{equation} \label{eq:unpert_vortcons}
\mathcal{L}_u \Delta \omega_{\alpha\beta} = 0 ,
\end{equation}
where $\Delta \omega_{\alpha\beta}$ denotes the Lagrangian
perturbation of the vorticity tensor (see Sec.~7.4.2 of
Ref.~\cite{Rot_rel_stars} for a derivation).

We suppose that the tidally deformed, slowly rotating star began in an
unperturbed state at $t=0$, and that the tidal field was switched on
adiabatically. Equation~\eqref{eq:unpert_vortcons} then implies that
\begin{equation} \label{vortcons}
\Delta \omega_{\alpha\beta} = 0
\end{equation}
for all time. This is the natural state that arises in a binary system
that was widely separated in the distant past. The \emph{vorticity
  preservation condition}~\eqref{vortcons} almost completely
constrains the perturbed fluid variables. Apart from the freedom to
add stationary $r$- and $g$-modes, which we suppress following
Refs.~\cite{Dynamical_response,Internal_metric}, there remains some
freedom in $u_A$. This freedom is fixed by the Einstein equation, and
all the perturbed fluid variables can be related to the radial
functions appearing in Eq.~\eqref{eq:interior_metric}.

We impose Eq.~\eqref{vortcons} to place the NS in a vorticity-preserving state, which we call the \emph{irrotational state} because its vorticity vanishes in the $\Omega \to 0$ limit. After a lengthy
calculation~\cite{Dynamical_response,Internal_metric}, the full
expressions for the perturbed fluid variables are found to be
\begin{subequations}
  \label{eq:perturbed_fluid}
  \begin{align} \label{emansatz}
\mu &= \bar{\mu} +  \frac{1}{2}e^{-2\psi}(\bar{\mu}+\bar{p}) \frac{d\bar{\mu}}{d\bar{p}} e^{\text{q}}_{tt} \mathcal{E}^{\text{q}} , \\
p &= \, \bar{p} + \frac{1}{2}e^{-2\psi}(\bar{\mu}+\bar{p}) e^{\text{q}}_{tt} \mathcal{E}^{\text{q}} , \\
u_r &= e^{-\psi} \hat{e}^{\text{q}}_{tr} \hat{\mathcal{E}}^{\text{q}} - t e^{-\psi}\left[\frac{1}{8\pi r^2 (\bar{\mu}+\bar{p})} - 1 \right] k^{\text{d}}_{tr1} \mathcal{K}^{\text{d}} - t e^{-\psi}\left[\frac{3}{4\pi r^2 (\bar{\mu}+\bar{p})} - 1 \right] k^{\text{o}}_{tr1} \mathcal{K}^{\text{o}} , \\
u_A &= -e^{-\psi} r^2 \omega \Omega^{\text{d}}_A -\frac{1}{6}r^2e^{-3 \psi } \left(1-\omega+\frac{d\bar{\mu}}{d\bar{p}} \right) e_{tt}^{\text{q}} \hat{\mathcal{E}}^{\text{q}}_A + e^{-\psi} f^{\text{d}}_t \mathcal{F}^{\text{d}}_A + e^{-\psi} f^{\text{o}}_t \mathcal{F}^{\text{o}}_A \nonumber \\ & \;\;\;\; + \frac{1}{3}t e^{-\psi} \omega b^{\text{q}}_t \hat{\mathcal{B}}^{\text{q}}_A - \frac{te^{-\psi}}{16 \pi r^2(\bar{\mu}+\bar{p})} \left\lbrace r^2 f \frac{d k^{\text{d}}_{tr1}}{dr} + 2 \left[ m - 2\pi r^3 (\bar{\mu}-\bar{p}) \right] k^{\text{d}}_{tr1} \right\rbrace \mathcal{K}^{\text{d}}_A  \nonumber\\ & \;\;\;\; - \frac{3te^{-\psi}}{16 \pi r^2(\bar{\mu}+\bar{p})} \left\lbrace r^2 f \frac{d k^{\text{o}}_{tr1}}{dr} + 2 \left[ m - 2\pi r^3 (\bar{\mu}-\bar{p}) \right] k^{\text{o}}_{tr1} \right\rbrace \mathcal{K}^{\text{o}}_A\label{emansatz3} .
  \end{align}
\end{subequations}
The time dependent terms in the fluid velocity represent dynamical currents induced by the stationary gravitomagnetic tidal field; they are tied to the time dependent radial functions introduced in Sec.~\ref{SubSec:Perturbed_metric}. The linear dependence on $t$ may be viewed as a consequence of our slow rotation approximation.

\subsection{Interior solution}\label{SubSec:Interior_solution}

With the ansatzes~\eqref{eq:interior_metric}
and~\eqref{eq:perturbed_fluid} for the metric and fluid variables, the
perturbed Einstein equation is solved in the stellar interior. The undetermined radial functions in the metric satisfy generically
inhomogeneous second-order ODEs. Thus, there exist two independent
homogeneous solutions, plus a particular solution, for each
differential equation. We demand that the solution be regular at the
origin, and that the interior solution match the exterior one at the
surface, up to first derivatives. While the system may appear to be
overdetermined at first glance, with three conditions on two free
parameters, we recall that the exterior solution also has a free
parameter: the Love number (or the moment of inertia in the case of
$\omega$). The matching condition fixes this final parameter, and
determines the tidal response.

In general, the interior ODEs must be integrated numerically, as they
depend on the fluid EoS, and we implement a shooting method to obtain
the solutions. We perform a local analysis of each ODE near $r=0$ to
determine the regularity conditions, and we then integrate outwards to
the surface. Here, the matching conditions determine the amplitude of
the regular solution and the free parameter of the exterior solution.

Consider, for example, the rotation. The function $\omega(r)$
satisfies
\begin{equation}\label{eq:2_5_Phil_paper}
  r f \frac{d^2 \omega}{dr^2} + \left[4 f - 4 \pi r^2 (\bar{\mu}+\bar{p}) \right]  \frac{d\omega}{dr} - 16 \pi r (\bar{\mu}+\bar{p}) \omega = 0
\end{equation}
inside the star. At the surface, $\omega$ matches on to the external
solution $\omega(r > R) = 1 - 2I/r^3$. Writing $' \equiv d/dr$,
regularity of $\omega$ at the origin requires $\omega'(0) =0$, with
$\omega(0)$ set by matching to the external solution. The matching
conditions also determine the specific value of $I$ that appears in
the exterior metric.

In the remainder of this section, we detail our method for calculating
the four scaled Love numbers
$\{
K_2^\text{el},K_2^\text{mag},\mathfrak{F}^{\text{o}},\mathfrak{K}^{\text{o}}
\}$. Like in the external problem, the relevant ODEs involve only a
small number of radial functions. In addition to the set
$\{e^{\text{q}}_{tt}, b^{\text{q}}_t, k^{\text{o}}_{tt},
f^{\text{o}}_t\}$ from above, we also require $k_{tr1}^{\text{o}}$,
which is non-vanishing in the interior and appears with
$k^{\text{o}}_{tt}$ in a coupled system of differential equations.

\subsubsection{Gravitoelectric sector: $K_2^{\text{el}}$ and $\mathfrak{F}^{\text{o}}$}\label{SubSec:Gravitoelectric}
The scaled gravitoelectric tidal Love number $K_2^{\text{el}}$ is determined by
solving for the radial function $e_{tt}^{\text{q}}$, which appears in
$g_{tt}$. In the interior of the star, this function satisfies the ODE
\cite{Internal_metric}
\begin{equation}\label{eq:2_9_Phil_paper}
  r^2 f \frac{d^2 e_{tt}^\text{q}}{dr^2} - 2 \left[ \frac{3m}{r} - 1 + 2 \pi r^2 (\bar{\mu}+3\bar{p}) \right] r \frac{d e_{tt}^\text{q}}{dr} - 2 \left[ 3 - 2 \pi r^2 (\bar{\mu}+\bar{p}) \left(3 + \frac{d \bar{\mu}}{d\bar{p}} \right)  \right] e_{tt}^\text{q} = 0.
\end{equation}
We would like to find a regular solution to this equation that matches the corresponding external expression given in Table~\ref{tb:external_metric} up to first derivatives at $r=R$. Local analysis of
Eq.~\eqref{eq:2_9_Phil_paper} near $r=0$ shows that the regular solution
has $e_{tt}^{\text{q}}(0) = e_{tt}^{\text{q}}{}'(0) = 0$, so its amplitude is determined by $e_{tt}^{\text{q}}{}''(0)$. The matching conditions at the surface then yield the value of $K^{\text{el}}_2$.

The scaled gravitoelectric rotational-tidal Love number $\mathfrak{F}^{\text{o}}$ is calculated from the solution for $f_t^{\text{o}}$, which satisfies \cite{Internal_metric}
\begin{gather}\begin{aligned}[b]
	\label{eq:4_8_Phil_paper}
	0 = & ~r^2 f \frac{d^2 f_t^{\text{o}}}{dr^2} - 4 \pi r^3 (\bar{\mu}+\bar{p}) \frac{d f_t^{\text{o}}}{dr} + 4 \left[ \frac{m}{r} - 3 + 2 \pi r^2 (\bar{\mu}+\bar{p}) \right] f_t^{\text{o}} \\
	 &-  16 \pi r^2 (\bar{\mu}+\bar{p}) f_t^{\text{o}} + r^3 e^{-2 \psi} (1 - \omega) \left( \frac{9m}{r} - 2 + 20 \pi r^2 \bar{p} \right) \frac{d e_{tt}^\text{q}}{dr} \\
	&+  2 r^2 e^{-2 \psi} \left[ \left(\frac{5 m}{r} + 2 \right)(1 - \omega) + 2 \pi r^2(\bar{\mu}+\bar{p}) \left(6 + \frac{d\bar{\mu}}{d\bar{p}}\right) \omega + 2 \pi r^2(\bar{\mu}+\bar{p}) \left(\frac{d\bar{\mu}}{d\bar{p}} - 2 \right)\right] e_{tt}^\text{q} 
	\end{aligned}
\end{gather}
in the interior. At the center, Eq.~\eqref{eq:4_8_Phil_paper} yields a
regular solution $f_t^{\text{o}} \propto r^4$. The matching conditions
at the surface then determine $\mathfrak{F}^\text{o}$.\footnote{As
  mentioned in Ref.~\cite{Internal_metric}, the value of
  $\mathfrak{F}^{\text{o}}$ is sensitive to the presence of stellar
  $r$-modes; such modes would modify
  Eq.~\eqref{eq:4_8_Phil_paper}. The values of
  $\mathfrak{F}^{\text{o}}$ computed here correspond to stars free of
  $r$-modes.}

\subsubsection{Gravitomagnetic sector: $K_2^{\text{mag}}$ and $\mathfrak{K}^{\text{o}}$}\label{SubSec:Gravitomagnetic}
The scaled Love numbers $K_2^{\text{mag}}$ and $\mathfrak{K}^{\text{o}}$ are
calculated in a similar fashion as $K_2^{\text{el}}$ and
$\mathfrak{F}^{\text{o}}$. The scaled gravitomagnetic tidal Love number is determined by solving for the radial function
$b_t^{\text{q}}$, which satisfies \cite{Dynamical_response}
\begin{equation}
  \label{eq:5_4_Phil_paper}
  r^2 f \frac{d^2 b_t^{\text{q}}}{dr^2} - 4 \pi r^3 (\bar{\mu}+\bar{p}) \frac{d b_t^{\text{q}}}{dr} - 2 \bigg [3 - \frac{2m}{r} - 4 \pi r^2 (\bar{\mu}+\bar{p}) \bigg ] b_t^\text{q} = 0.
\end{equation}
The regularity condition derived from a local analysis of
\eqref{eq:5_4_Phil_paper} is $b_t^{\text{q}} \propto r^3$ near
$r=0$. The matching procedure at $r=R$ yields the value of
$K_2^{\text{mag}}$.

The calculation of $\mathfrak{K}^{\text{o}}$ is slightly more
complicated, since two coupled ODEs must be solved. First, we determine the radial function
$k_{tr1}^{\text{o}}$ throughout the star. It
satisfies the ODE \cite{Dynamical_response}
\begin{gather}
  \begin{aligned}[b]
    \label{eq:5_15_Phil_paper}
    0 = & ~ r^2 f \frac{d^2 k_{tr1}^{\text{o}}}{dr^2} + \bigg [3(m - 4 \pi r^3 \bar{\mu}) + (m + 4 \pi r^3 \bar{p}) \frac{d\bar{\mu}}{d\bar{p}} \bigg ] \frac{d k_{tr1}^{\text{o}}}{dr} \\
    &- \frac{2}{r^2 f} \bigg \{ 2 \left[ 3 - 5 \pi r^2( \bar{\mu}+\bar{p}) + 8 \pi^2 r^4 \bar{p}^2 \right] r^2 - 2 \left[5 - 2 \pi r^2( 5 \bar{\mu} + 7 \bar{p}) \right] rm - 3 m^2 - \left( m + 4 \pi r^3 \bar{p} \right)^2 \frac{d \bar{\mu}}{d\bar{p}} \bigg \} k_{tr1}^{\text{o}} \\
    &- \frac{32 \pi}{3} r^2 (\bar{\mu}+\bar{p}) \omega \frac{d b_t^{\text{q}}}{dr} + \frac{16 \pi}{3} (\bar{\mu}+\bar{p}) \bigg [ r^2 \frac{d \omega}{dr} + 2 \frac{3 r - 7m - 4 \pi r^3 \bar{p}}{f} \omega \bigg ] b_t^{\text{q}}.
  \end{aligned}
\end{gather}
Since $k_{tr1}^{\text{o}}$ vanishes outside the star, the matching
conditions at the surface are simply
$k_{tr1}^{\text{o}}(R) = k_{tr1}^{\text{o}}{}' (R) = 0$. (Note that
these boundary conditions do not overdetermine the
system~\cite{Dynamical_response}.) Based on a local analysis of
Eq.~\eqref{eq:5_15_Phil_paper}, the regular solution has
$k_{tr1}^{\text{o}} \propto r^4$ at the origin.

Second, we solve for the radial function $k_{tt}^{\text{o}}$, which satisfies \cite{Dynamical_response}
\begin{gather}
	\begin{aligned}[b]
	\label{eq:5_16_Phil_paper}
	0 = & ~ r^2 f \frac{d^2 k_{tt}^{\text{o}}}{dr^2} + 2 \bigg [1 - \frac{3m}{r} - 2 \pi r^2 (\bar{\mu}+3\bar{p}) \bigg ] r \frac{d k_{tt}^{\text{o}}}{dr} + 4 \bigg [\pi r^2 (\bar{\mu}+\bar{p}) \left( 3 + \frac{d \bar{\mu}}{d\bar{p}} \right) -3 \bigg ] k_{tt}^{\text{o}} + \frac{1}{2} r^2 f \bigg (\frac{d \bar{\mu}}{d\bar{p}} -1 \bigg ) \frac{d k_{tr1}^{\text{o}}}{dr} \\
	&+ \left \{ \left (11 + \frac{d \bar{\mu}}{d\bar{p}} \right ) m + 2 \pi r^3 \left [ (\bar{\mu}+7\bar{p}) - (\bar{\mu}-\bar{p}) \frac{ d \bar{\mu}}{d\bar{p}} \right ] - 4r \right \} k_{tr1}^{\text{o}} + S_1 r \frac{d b_t^{\text{q}}}{dr} + S_0 b_t^{\text{q}} ,
	\end{aligned}
\end{gather}
with
\begin{subequations}
  \begin{alignat}{2}
    \label{eq:5_17_Phil_paper}
    S_1 = & - \frac{2}{3} \bigg \{ \frac{5 m^2}{r^2} + 3 - 16 \pi^2 r^4 \bar{p}^2 + 4 \pi r^2 \bar{\mu} - \frac{m}{r} \left[ 9 + 8\pi r^2 (\bar{\mu}+\bar{p}) \right] \bigg \} r \frac{d \omega}{dr} \nonumber \\
    & -\frac{4}{3} \bigg \{ 3 - \frac{m}{r} \left[ 9 - 4 \pi r^2 (\bar{\mu}+\bar{p}) \right] - 4 \pi r^2 \bar{p} \left[ 3 - 4 \pi r^2 (\bar{\mu}+\bar{p}) \right] \bigg \} \omega + 4 \bigg (1 - \frac{3m}{r} - 4 \pi r^2 \bar{p} \bigg ), \\
    S_0 = & ~ \frac{2}{3} \bigg \{ \frac{10 m^2}{r^2} + 4 \pi r^2 \left[ \left( 3 - 8 \pi r^2 \bar{p} \right)\bar{p} + 2 \bar{\mu} \right] - \frac{m}{r} \left[ 3+ 16 \pi r^2 (\bar{\mu}+\bar{p}) \right] \bigg \} r \frac{d \omega}{dr} \nonumber \\
    & + \frac{4}{3} \bigg \{ \frac{m}{r} \left[ 6 + 8 \pi r^2 (\bar{\mu}+\bar{p}) \right] + \bigg [9 - 4 \pi r^2 (\bar{\mu}+\bar{p}) \bigg (6 + \frac{d \bar{\mu}}{d\bar{p}} - 8 \pi r^2 \bar{p} \bigg ) \bigg ] \bigg \} \omega \nonumber \\
    & - 4 \bigg \{ \frac{2m}{r} - \bigg [ 2 \pi r^2 (\bar{\mu}+\bar{p}) \bigg (1 + \frac{d \bar{\mu}}{d\bar{p}} \bigg ) - 3 \bigg ] \bigg \} .
  \end{alignat}
\end{subequations}
A local analysis of Eq.~\eqref{eq:5_16_Phil_paper} reveals that the
regular solution has $k_{tt}^{\text{o}} \propto r^3$ near the
origin. Finally, the matching conditions at the surface determine
$\mathfrak{K}^{\text{o}}$.

\subsection{Redefinition of $\mathfrak{F}^\text{o}$}\label{SubSec:f_o}
As noted in the introduction, the definition of the scaled gravitoelectric
rotational-tidal Love number $\mathfrak{F}^\text{o}$ given in this
work differs from that of Landry and
Poisson~\cite{External_metric,Internal_metric}; namely,
\begin{equation}
  \label{eq:def_f_o_new}
  \mathfrak{F}^\text{o}[\text{here}] = \mathfrak{F}^\text{o}[\text{LP}] - \frac{5}{3}K_2^\text{el}, \qquad \text{which implies} \qquad \mathfrak{f}^\text{o}[\text{here}] = \mathfrak{f}^\text{o}[\text{LP}] + \frac{5}{3}k_2^\text{el}.
\end{equation}
We claim that our new definition is more consistent with the
interpretation of $\mathfrak{F}^{\text{o}}$ as a scaled rotational-tidal Love
number.

\begin{figure}[t]
\includegraphics[]{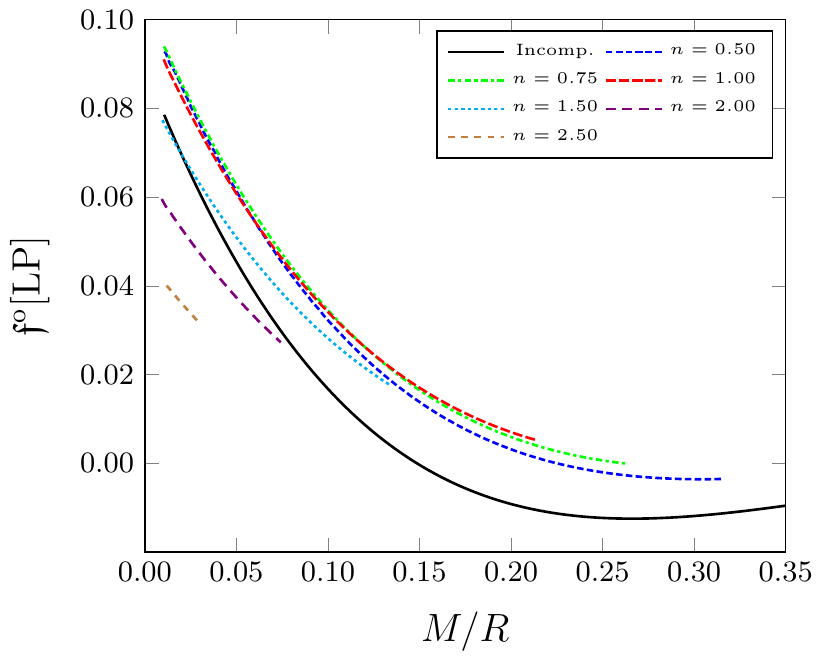}
\caption{Gravitoelectric rotational-tidal Love number
  $\mathfrak{f}^\text{o}[\text{LP}]$ as a function of compactness. This is a
  reproduction of Fig.~1 of Ref.~\cite{Internal_metric} with the addition
  of the incompressible fluid results. We employ a different definition of $\mathfrak{f}^{\text{o}}$ elsewhere in the paper---see Eq.~\eqref{eq:def_f_o_new}.}
\label{fig:f_o_old}
\end{figure}

With Landry and Poisson's definition of $\mathfrak{F}^{\text{o}}$, the
octupole part of the $tA$ component of the exterior metric takes the
schematic form
\begin{equation} \label{octgta}
g_{tA}^{\ell = 3} = - \frac{10 M}{3r^2} \left[ \frac{M^2}{r} + \ldots + \frac{6}{5} \mathfrak{f}^{\text{o}}[\text{LP}] \left(\frac{R}{r}\right)^5 \left( 1 + \ldots \right) + 2 k^{\text{el}}_2 \left(\frac{R}{r}\right)^5 \left( 1 + \ldots \right) \right] \epsilon_{ac}^{\;\;\;\,d} x^c \mathcal{E}_{\langle db} \chi_{c \rangle} x^b x^e x^a_A ,
\end{equation}
rather than that of Eq.~\eqref{gttb}. (This expression can be obtained
by expanding the radial function $f_t^{\text{o}}$ from Table IV of
Ref.~\cite{External_metric} in powers of $M/r$.)  We see that there
are two separate pieces that decay in $r$ at the same rate, which
means that $\mathfrak{F}^{\text{o}}[\text{LP}]$ is only
\emph{partially} measuring the response of the body to the
spin-coupled gravitoelectric field. The fact that the
$\mathfrak{F}^{\text{o}}[\text{LP}]$ and $K_2^{\text{el}}$ terms in
Eq.~\eqref{octgta} have the same scaling with $r$, however, allows us
to shift the rotational-tidal Love number as in
Eq.~\eqref{eq:def_f_o_new} so that it fully captures the response.

The unnaturalness of Landry and Poisson's definition is clearly
demonstrated in Fig.~\ref{fig:f_o_old}, reproduced from
Ref.~\cite{Internal_metric} with the addition of the incompressible
fluid results. We see that the polytrope
$\mathfrak{f}^{\text{o}}[\text{LP}]$ vs.~compactness curves intersect
one another and do not tend monotonically to the incompressible
fluid. This is contrary to physical intuition, and it differs
qualitatively from results for the other three Love numbers (see
Fig.~\ref{fig:Incompressible}). These discrepancies disappear when our
new definition for the rotational-tidal Love number is used, as
Fig.~\ref{subfig:fo} shows.

We have two additional comments on our definition of
$\mathfrak{F}^\text{o}$. First, the shift in
Eq.~(\ref{eq:def_f_o_new}) has the same effect in the light-cone
gauge~\cite{Preston} employed in Ref.~\cite{External_metric} as it
does here. This confirms that our new definition is not just a quirk
of the Regge-Wheeler gauge. Second, our new definition of
$\mathfrak{F}^\text{o}$ coincides with the scaled rotational-tidal Love
number $\delta \tilde{\lambda}_M^{(32)}$ of Ref.~\cite{Pani} up to a
constant factor~\cite{Internal_metric}.

\section{Love numbers for polytropes}\label{Sec:results}

In this section, we test our methods by computing Love numbers for
polytropes. We show that our results are consistent with similar
calculations in the literature, and that when we take suitable limits
they match post-Newtonian and incompressible fluid calculations. The
consistency checks we perform are important as they provide support
for our findings, which disagree with those of Ref.~\cite{Pani}.

\begin{figure}
\centering
\begin{subfigure}[b]{0.45\linewidth}
\includegraphics[]{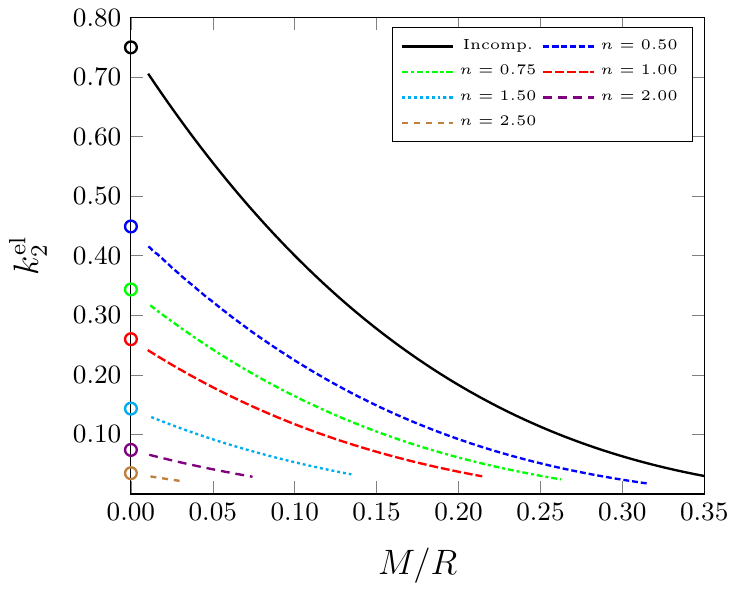} 
\caption{Gravitoelectric tidal Love number}
\end{subfigure}
\begin{subfigure}[b]{0.45\linewidth}
\includegraphics[]{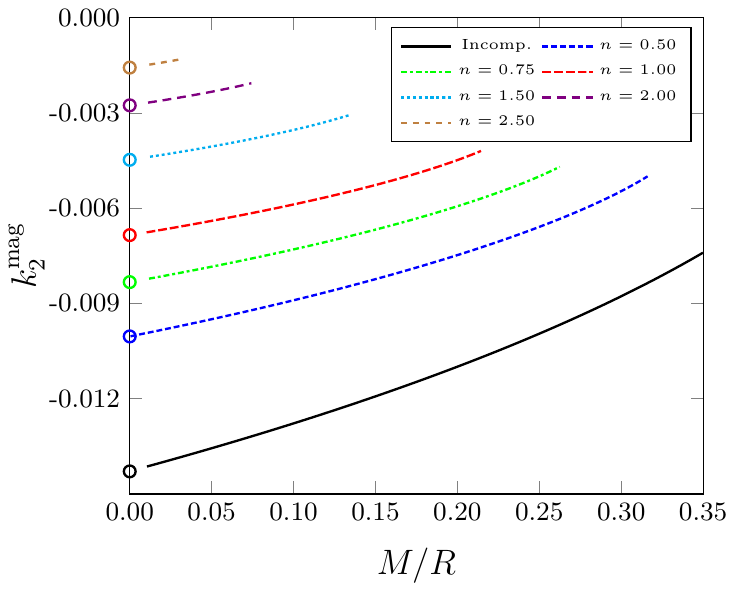} 
\caption{Gravitomagnetic tidal Love number}
\end{subfigure}
\begin{subfigure}[b]{0.45\linewidth}
\includegraphics[]{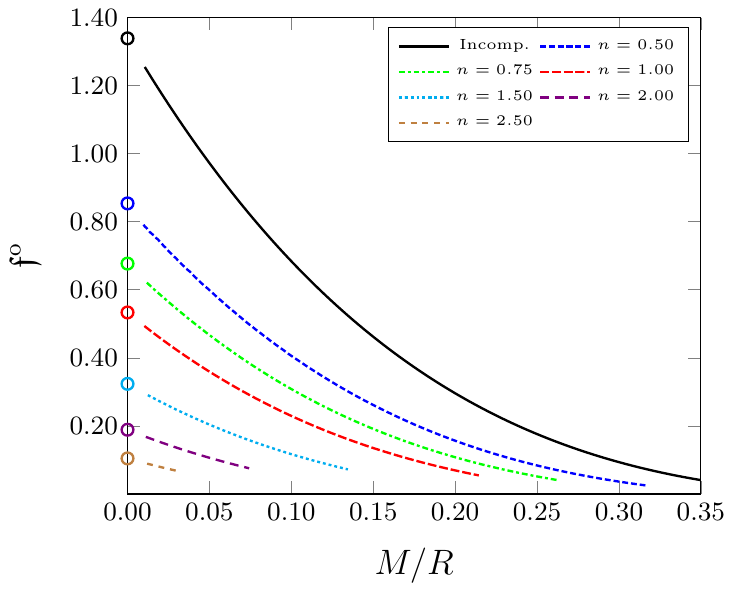} 
\caption{Gravitoelectric rotational-tidal Love number} \label{subfig:fo}
\end{subfigure}
\begin{subfigure}[b]{0.45\linewidth}
\includegraphics[]{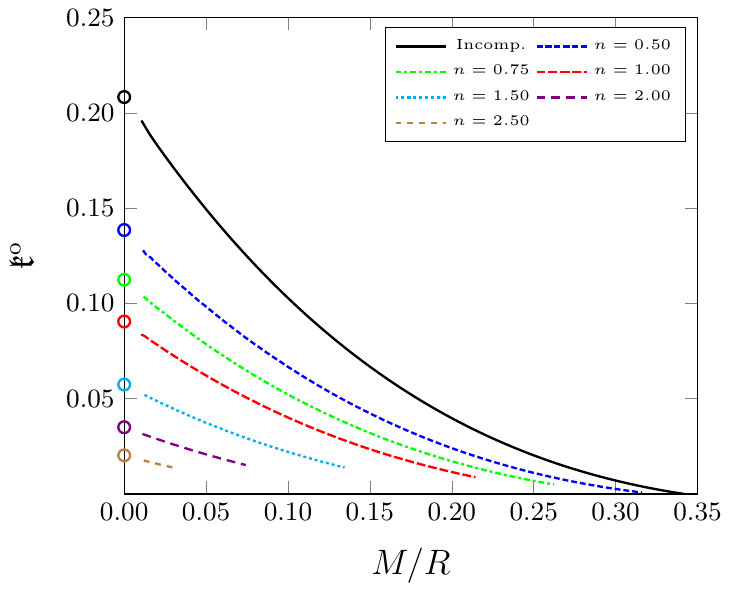} 
\caption{Gravitomagnetic rotational-tidal Love number}
\end{subfigure}
\caption{Love numbers
  $\{ k_2^\text{el}, k_2^\text{mag}, \mathfrak{f}^\text{o},
  \mathfrak{k}^\text{o} \}$ as a function of compactness $M/R$ for
  different polytropes and an incompressible fluid star. Circles
  indicate Love numbers calculated in the post-Newtonian approximation
  (see also Table~\ref{tb:PN_LN}).}
\label{fig:Incompressible}
\end{figure}

In Fig.~\ref{fig:Incompressible} we plot the Love numbers for various
polytropes as a function of stellar compactness. We use the mass
polytrope EoS $p(\rho) = K\rho^{1 + 1/n}$, where $n>0$ is the
\emph{polytropic index} and $K$ is a constant.\footnote{Here, $p$ and
  $\rho$ refer to the \emph{background} pressure and rest mass
  density. From this section on, we deal only with unperturbed fluid
  quantities, and we therefore drop the overbars on the background
  $\mu$ and $p$.} A choice of $n$ and central pressure-to-density
ratio $p_c/\rho_c = K \rho_c^{1/n}$ uniquely specifies the stellar
model. The Love numbers are plotted up to the maximum value of $M/R$
for which the given polytrope is stable against radial perturbations,
as per the turning point criterion (see e.g.,
Ref.~\cite{Rot_rel_stars}). Our tidal Love number results agree with
Refs.~\cite{Binnington} and~\cite{Irrotational}, while our
rotational-tidal Love numbers match those of
Ref.~\cite{Internal_metric} (modulo the redefinition of
$\mathfrak{F}^{\text{o}}$).

Figure \ref{fig:Incompressible} also includes Love numbers for
incompressible fluid stars. The incompressible fluid constitutes the
(singular) $n \to 0$ limit of the polytropic EoS. We see that the
polytrope curves tend monotonically toward that of the incompressible
fluid as $n$ decreases. The fact that the incompressible fluid
possesses the largest Love numbers (in magnitude) of all the models is
consistent with the physical intuition that, as the stiffest possible
EoS, it should have the weakest internal gravity and hence the
greatest deformability. Subtleties associated with the
calculation of Love numbers for this uniform-density model are treated
in Appendix~\ref{SubSec:Incompress}.

We also show polytrope and incompressible fluid Love numbers computed
in post-Newtonian theory (indicated by circles on the vertical
axes). Details of these \emph{independent} calculations are provided
in Appendix~\ref{SubSec:PN_calc} and the results are listed in
Table~\ref{tb:PN_LN}. It is apparent from
Fig.~\ref{fig:Incompressible} that the general-relativistic Love
numbers match the post-Newtonian values in the weak-field limit
$M/R \to 0$, as expected. This agreement further validates our Love
number computations.

\begin{table}
\begin{ruledtabular}
	\centering
	\begin{tabular}{ M{0.2\textwidth} M{0.2\textwidth} M{0.2\textwidth} M{0.2\textwidth} M{0.2\textwidth} }
	EoS & $k_2^\text{el}$ & $k_2^\text{mag}~(\times10^{-3})$ & $\mathfrak{f}^{\text{o}}$ & $\mathfrak{k}^{\text{o}}$\\ 
	\hline \noalign{\medskip}
	Incompressible & 0.7500\phantom{5} & -14.29 & 1.339\phantom{5} & 0.2083\phantom{5}  \\
	n = 0.50 & 0.4492\phantom{5} & -10.04 & 0.8539 & 0.1385\phantom{5}  \\
	n = 0.75 & 0.3434\phantom{5} & -8.330 & 0.6770 & 0.1124\phantom{5}  \\
	n = 1.00 & 0.2599\phantom{5} & -6.850 & 0.5334 & 0.09050  \\
	n = 1.50 & 0.1433\phantom{5} & -4.478 & 0.3235 & 0.05740  \\
	n = 2.00 & 0.07394 & -2.763 & 0.1886 & 0.03501  \\
	n = 2.50 & 0.03485 & -1.575 & 0.1044 & 0.02026  \\
	\end{tabular}
\end{ruledtabular}
\caption{Post-Newtonian values of the Love numbers for the EoSs considered
  in Fig.~\ref{fig:Incompressible}, calculated with the methods of
  Appendix~\ref{SubSec:PN_calc}.}
	\label{tb:PN_LN}
\end{table}

\begin{figure}
\includegraphics[]{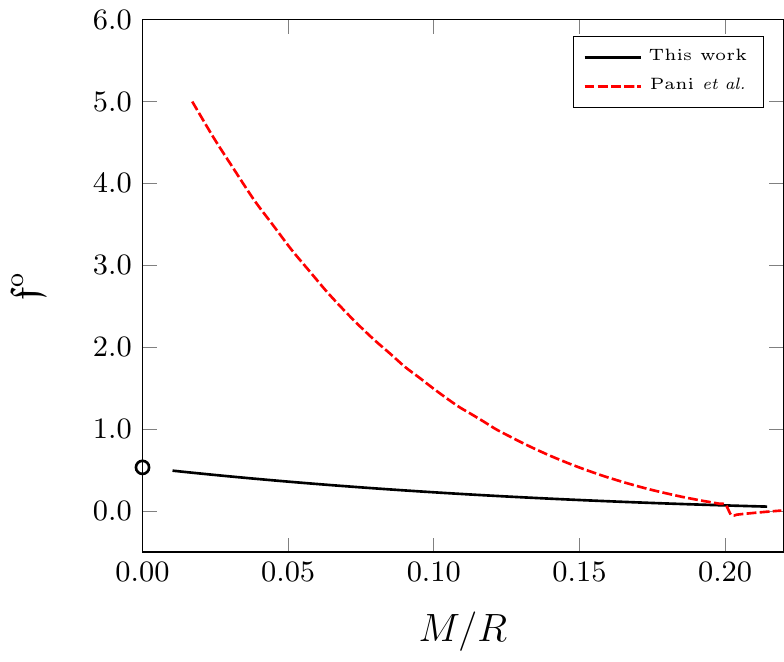} 
\caption{\label{fig:Pani}Gravitoelectric rotational-tidal
  Love number $\mathfrak{f}^\text{o}$ for an $n=1$ polytrope. The
  solid black curve corresponds to our mass-polytrope results, whereas the
  dashed red curve is inferred via Eq.~\eqref{eq:map} from the energy-polytrope results labelled ``POLYn1" in the bottom right panel of Fig.~5 of Ref.~\cite{Pani}. The black circle shows the result of the
  post-Newtonian calculation, which both mass and energy polytropes
  should match.}
\end{figure}

We now make a direct comparison of our results for the
rotational-tidal Love numbers with those of Ref.~\cite{Pani}. Pani, Gualtieri and Ferrari employ slightly different definitions for the Love numbers, but we derive the relation between the two conventions in Appendix~\ref{SubSec:PGF}. The scaled gravitoelectric and gravitomagnetic rotational-tidal Love numbers of Ref.~\cite{Pani}, respectively $\delta \tilde{\lambda}^{(32)}_M$ and $\delta \tilde{\lambda}^{(32)}_E$, are merely rescaled by a constant factor relative to our own definitions:

\begin{equation} \label{eq:map}
\delta \tilde{\lambda}^{(32)}_M = \frac{96}{\sqrt{5\pi}} \mathfrak{F}^{\text{o}} , \qquad \delta \tilde{\lambda}^{(32)}_E = -144 \sqrt{\frac{7}{5}} \mathfrak{K}^{\text{o}} .
\end{equation}
Provided our model of the NS's fluid interior is the same, our Love numbers can be compared with this mapping.
Whereas in general the two frameworks make different assumptions about the state
of the fluid---irrotational versus static---these states coincide in
the gravitoelectric octupole sector when the star is free of $r$- and
$g$-modes~\cite{Irrotational,Internal_metric}. Thus, for the same EoS,
both sets of results for $\mathfrak{F}^\text{o}$ should agree. Pani, Gualtieri and Ferrari computed Love numbers for an $n=1$ energy
polytrope with EoS $p(\mu) = K\mu^{1 + 1/n}$, which should coincide
with our mass polytrope in the weak-field limit $M/R \rightarrow
0$. Nevertheless, Fig.~\ref{fig:Pani} shows disagreement between our
results and those of Ref.~\cite{Pani}. The fact that our
general-relativistic results agree in the limit with the
post-Newtonian calculation of $\mathfrak{f}^\text{o}$ gives us
confidence in the conclusions we draw from our work, and
simultaneously raises some questions about the computations of
Ref.~\cite{Pani}, at least at low compactness.

\section{Love numbers for realistic equations of state}\label{Sec:RealEoS}

Having checked our method on polytropes, we now turn to realistic
EoSs. We use a piecewise polytrope approximation to the tabulated EoSs
computed in nuclear theory. In the stellar core, each model consists
of a three-piece polytrope with an overall scale, resulting in four
EoS-dependent parameters in total. The crust is always described by a
four-piece polytropic approximation to the SLy EoS. This description
of realistic EoSs follows closely that of Ref.~\cite{Read_etal09},
which we summarize here.

The stellar core EoS is taken to be
\begin{equation}\label{eq:piecewise_polytrope}
p(\rho) = \left\{ \begin{array}{ll} 
K_1 \rho^{\Gamma_1}, & \quad \rho_0 \leq \rho \leq \rho_1,\\
K_2 \rho^{\Gamma_2}, & \quad \rho_1 \leq \rho \leq \rho_2,\\
K_3 \rho^{\Gamma_3}, & \quad \rho_2 \leq \rho,\\
\end{array} \right.
\end{equation}
with the transition rest mass densities
$\rho_1 \equiv 10^{14.7}\text{g/cm}^{3}$ and
$\rho_2 \equiv 10^{15.0}\text{g/cm}^{3}$. (The transition density
$\rho_0$ is determined by matching at the crust-core interface.) The
parameters $K_i$ are determined in terms of the overall scale of
$p(\rho)$ by requiring continuity.  Indeed, evaluating $p(\rho_1)$
sets $K_1 = p(\rho_1) \rho_1^{-\Gamma_1}$, and imposing continuity at
the interfaces $\rho_1$ and $\rho_2$ gives the recursive formula
\begin{equation}
K_{i+1} = K_i \rho_i^{\Gamma_i - \Gamma_{i+1}}, \quad i=1,2.
\end{equation}
Thus, the four parameters
$\{ p(\rho_1), \Gamma_1, \Gamma_2, \Gamma_3 \}$ specify the EoS in the
stellar core. Using continuity, the internal energy density $\epsilon$
is determined by the first law~\eqref{eq:first_law} to be
\begin{equation}\label{eq:thermo_energy}
\epsilon(\rho; \rho_{i-1}\le \rho \le \rho_i ) = a_i \rho + \frac{1}{\Gamma_i - 1} K_i \rho^{\Gamma_i}, \qquad \text{with} \qquad a_i = \frac{\epsilon(\rho_{i-1})}{\rho_{i-1}} - \frac{K_i}{\Gamma_i - 1} {\rho}^{\Gamma_i - 1}_{i-1}, \qquad i=1,2,3,
\end{equation}
and from this, we obtain the the total energy density $\mu$. Since
$a_i$ depends on $\epsilon(\rho_{i-1})$, $a_1$ is determined from the
crust-core interface, and then $a_2$ and $a_3$ follow recursively.

The stellar crust model is similar to the core, except: (1) It
consists of four polytropic phases---indexed by \linebreak
$i=-3,-2,-1,0$---instead of three; (2) Instead of $p(\rho_{-3})$, we
are directly given the constants $\{ K_{-3}, K_{-2}, K_{-1}, K_0 \}$;
and (3) The crust model depends on the core via the interface density,
\begin{equation}\label{eq:rho_0}
\rho_0 = \left( \frac{K_1}{K_0} \right) ^{1/(\Gamma_0-\Gamma_1)}.
\end{equation}
The remaining parameter needed to describe the crust, $a_{-3}$, is
fixed to zero by requiring that $\epsilon/\rho \to 0$ in the zero rest
mass density limit. Numerical parameters for the SLy crust model are
provided in Ref.~\cite{Read_etal09}, and we display them in
Table~\ref{tb:Crust_parameters}.
\begin{table}[bt]
	\begin{ruledtabular}
	\centering
	\begin{tabular}{ M{0.25\textwidth} M{0.25\textwidth} M{0.25\textwidth}  M{0.25\textwidth} }
	Phase & $K_i$ & $\Gamma_i$ & $\rho_i ~ (\text{g/cm}^3)$ \\ 
	\hline \noalign{\medskip}
	 -3 & $6.11252 \times 10^{12}$ & 1.58425 & $2.44034 \times 10^{7\phantom{0}}$  \\ 
	 -2 & $9.54352 \times 10^{14}$ & 1.28733 & $3.78358 \times 10^{11}$ \\ 
	 -1 & $4.78764 \times 10^{22}$ & 0.62223 & $2.62780 \times 10^{12}$ \\ 
	  0 & $3.59389 \times 10^{13}$ & 1.35692 & $\rho_0$ \\ 
	\end{tabular}
	\end{ruledtabular}
	\caption{Parameters of the four-piece polytropic crust
          model. The constants $K_i$ are in cgs units so that
          $p=K_i \rho^{\Gamma_i}$ is in dyne/c$\text{m}^2$. The
          crust-core interface density $\rho_0$ depends on the core
          parameters $K_1$ and $\Gamma_1$ through Eq.~(\ref{eq:rho_0}).}
\label{tb:Crust_parameters}
\end{table}
Given the core, the crust is fully specified, so the
full stellar EoS is determined by the parameters
$\{ p(\rho_1), \Gamma_1, \Gamma_2, \Gamma_3 \}$.

Of the 34 candidate EoSs studied in Ref.~\cite{Read_etal09}, we
consider the seven that give rise to stable stars with maximum masses greater
than $2M_{\odot}$ while avoiding superluminal sound
propagation. This is consistent with the highest
observed NS masses of $(1.97~\pm~0.04)M_\odot$~\cite{Demorest} and
$(2.01~\pm~0.04)M_\odot$~\cite{Antoniadis}. Five of the EoSs describe pure
$npe\mu$ nuclear matter---SLy~\cite{fDpH01},
ENG~\cite{lEeOmHgBeO96}, MPA1~\cite{hMmPtA87}, MS1~\cite{hMbS96}, and
MS1b (which is identical to MS1 but with a low symmetry energy of $25$
MeV~\cite{Read_etal09}). The other two, H4~\cite{bLmNbO06} and
ALF2~\cite{mAmBmPsR04}, include nonstandard nuclear components
(hyperons and color-flavor-locked quark matter,
respectively). Parameters for the piecewise polytrope approximations
are given in Table~\ref{tb:EoS_parameters}.
\begin{table}[bt]
\begin{ruledtabular}
	\centering
	\begin{tabular}{ M{0.12\textwidth} M{0.12\textwidth} M{0.12\textwidth} M{0.12\textwidth} M{0.12\textwidth} M{0.12\textwidth} M{0.12\textwidth} M{0.12\textwidth} }
	EoS & SLy & ENG & MPA1 & MS1 & MS1b & H4 & ALF2 \\ 
	\hline \noalign{\medskip}
	$\log_{10} p(\rho_1)$  & 34.384 & 34.437 & 34.495 & 34.858 & 34.855 & 34.669 & 34.616  \\ 
	$\Gamma_1$  & \hspace{0.2cm}3.005 & \hspace{0.15cm}3.514 & \hspace{0.15cm}3.446 & \hspace{0.15cm}3.224 & \hspace{0.15cm}3.456 & \hspace{0.15cm}2.909 & \hspace{0.15cm}4.070 \\ 
	$\Gamma_2$  & \hspace{0.2cm}2.988 & \hspace{0.15cm}3.130 & \hspace{0.15cm}3.572 & \hspace{0.15cm}3.033 & \hspace{0.15cm}3.011 & \hspace{0.15cm}2.246 & \hspace{0.15cm}2.411 \\ 
	$\Gamma_3$  & \hspace{0.2cm}2.851 & \hspace{0.15cm}3.168 & \hspace{0.15cm}2.887 & \hspace{0.15cm}1.325 & \hspace{0.15cm}1.425 & \hspace{0.15cm}2.144 & \hspace{0.15cm}1.890 \\ 
	\end{tabular}
\end{ruledtabular}
\caption{Parameters for the piecewise polytropic fits modelling the
  EoSs that we consider. In this table, $p$ is measured in units of
  $\text{dyne}/\text{cm}^2$.}
\label{tb:EoS_parameters}
\end{table} 

We restrict our attention to configurations with masses of
astrophysical relevance
($M > M_{\odot}$~\cite{fOdPrNaS12,jMkSpFjDfJmMmBsBaR15}) that are
stable against radial perturbations according to the turning point
criterion.  In addition, the crust in realistic NSs should constitute
a small fraction of the whole star~\cite{Chamel_Haensel}. Defining
$r_{cc}$ as the radial position of the crust-core interface, and
$M_{cr} \equiv M - m(r_{cc})$ as the crust's mass, we note that our
configurations satisfy
\begin{subequations}\label{eq:crust_bounds}
\begin{eqnarray}
M_{cr} &<& 0.045M, \label{eq:crust_bound_mass}\\
R - r_{cc} &<& 0.15R. \label{eq:crust_bound_radius}
\end{eqnarray}
\end{subequations}

\begin{figure}
\centering
\begin{subfigure}[b]{0.45\linewidth}
\includegraphics[]{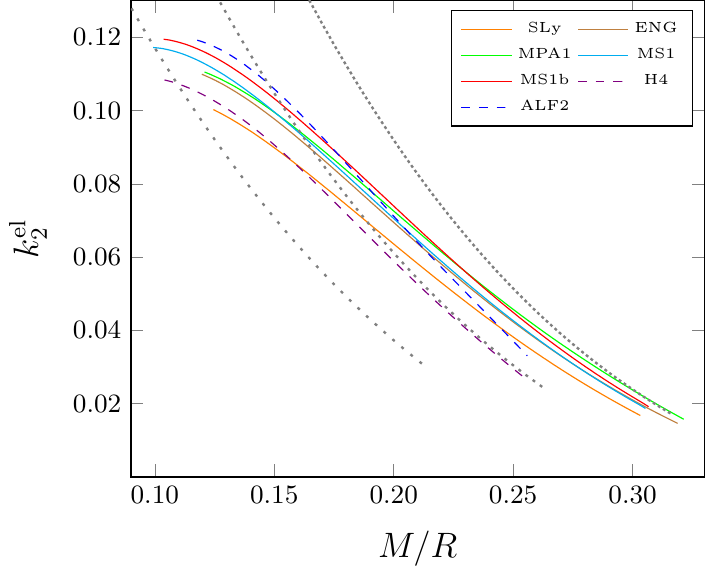} 
\caption{Gravitoelectric tidal Love number}
\end{subfigure}
\begin{subfigure}[b]{0.45\linewidth}
\includegraphics[]{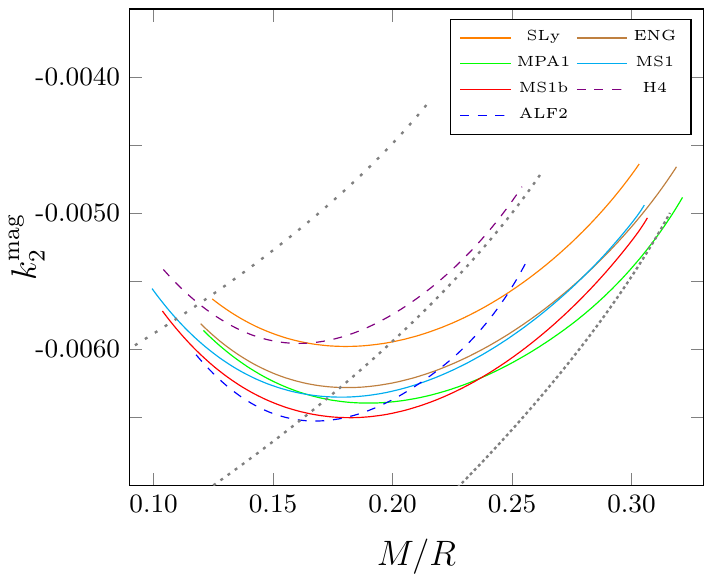} 
\caption{Gravitomagnetic tidal Love number}
\end{subfigure}
\begin{subfigure}[b]{0.45\linewidth}
\includegraphics[]{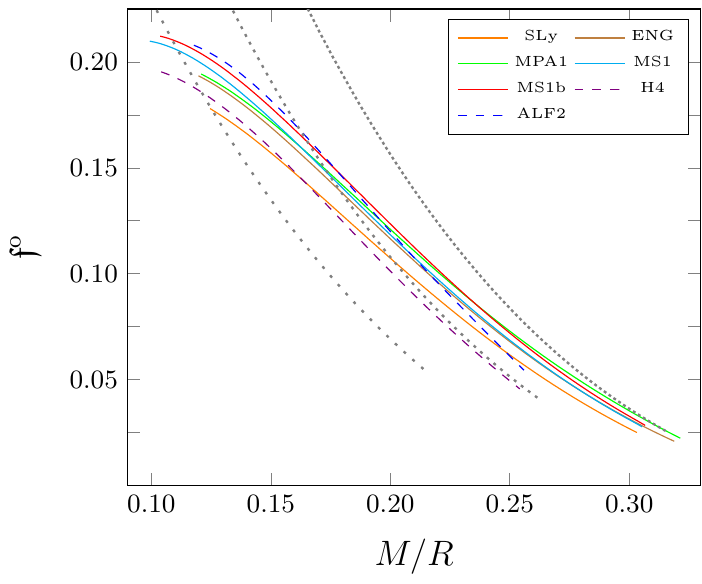} 
\caption{Gravitoelectric rotational-tidal Love number}
\end{subfigure}
\begin{subfigure}[b]{0.45\linewidth}
\includegraphics[]{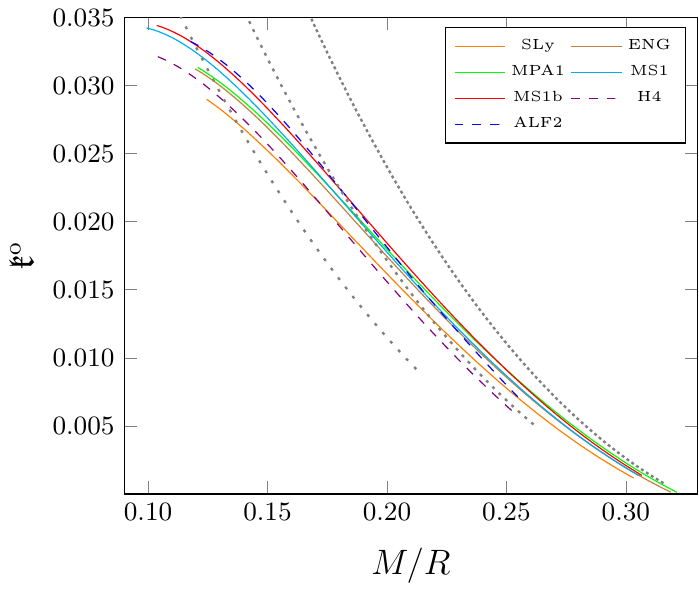} 
\caption{Gravitomagnetic rotational-tidal Love number}
\end{subfigure}
\caption{Love numbers for realistic-EoS NS models. Solid lines are used for $npe\mu$-matter EoSs, while dashed lines denote exotic-matter EoSs. For comparison, we plot results for $n=0.5$, $n=0.75$ and $n=1$  polytropes with dotted lines (with more densely-spaced dots corresponding to smaller $n$).}
\label{fig:Realistic_EoS}
\end{figure}

In Fig.~\ref{fig:Realistic_EoS} we plot the Love numbers
$\{ k_2^\text{el}, k_2^\text{mag}, \mathfrak{f}^{\text{o}},
\mathfrak{k}^{\text{o}} \}$ for our seven EoSs as functions of the
stellar compactness. For comparison, we include Love numbers for
$n=0.5$, $n=0.75$, and $n=1$ polytropes. We also display the Love
numbers corresponding to a canonical $1.4M_{\odot}$ NS in
Table~\ref{tb:LNs}.
\begin{table}
\begin{ruledtabular}
	\centering
	\begin{tabular}{ M{0.15\textwidth} M{0.15\textwidth} M{0.15\textwidth} M{0.15\textwidth} M{0.15\textwidth} M{0.15\textwidth} }
	EoS & $M/R$ & $k_2^\text{el}$ & $k_2^\text{mag} (\times 10^{-3})$ & $\mathfrak{f}^{\text{o}}$ & $\mathfrak{k}^{\text{o}}(\times 10^{-2})$\\ 
	\hline \noalign{\medskip}
SLy	& $0.1766$ & $0.07617$  & $-5.977$ & $0.1308$ & $2.048$ \\
ENG	& $0.1727$ & $0.08530$ & $-6.273$ & $0.1457$ & $2.272$\\
MPA1 & $0.1662$ & $0.09130$ & $-6.341$ & $0.1559$ & $2.435$ \\
MS1	& $0.1385$ & $0.10544$ & $-6.179$ & $0.1845$ & $2.967$ \\
MS1b & $0.1423$ & $0.10714$ & $-6.328$ & $0.1860$ & $2.968$ \\
H4 & $0.1482$ & $0.09162$ & $-5.930$ & $0.1611$ & $2.604$ \\
ALF2 & $0.1627$ & $0.09777$ & $-6.522$ & $0.1672$ & $2.619$ \\
		\end{tabular}
\end{ruledtabular}
	\caption{Love numbers corresponding to a $1.4M_{\odot}$ NS for each realistic EoS studied in Sec.~\ref{Sec:RealEoS}. The star's compactness $M/R$ is also listed.}
	\label{tb:LNs}
\end{table} 

A number of interesting features are apparent in
Fig.~\ref{fig:Realistic_EoS}. First, all the Love numbers have a
definite sign; we do not observe the zero-crossings reported in
Ref.~\cite{Pani}. Second, the Love numbers for realistic EoSs are
clustered between the $n=0.5$ and $n=1$ polytrope curves. There is,
however, a qualitative difference between the slope of the curves for
the $npe\mu$-matter EoSs (SLy, ENG, MPA1, MS1, MS1b; shown solid in
Fig.~\ref{fig:Realistic_EoS}) and the remaining two (H4, ALF2; shown
dashed). Love numbers for the $npe\mu$-matter EoSs also approach the
$n=0.5$ polytrope curve at high compactness. This limiting behaviour
is expected for SLy, ENG and MPA1, since---according to
Table~\ref{tb:EoS_parameters}---those EoSs have adiabatic indices
$\Gamma_3$ close to 3 (i.e. $n \approx 0.5$).

The limiting behavior also demonstrates that the crust makes
a negligible contribution to the Love numbers at large compactness,
since its adabiatic indices $\Gamma_i$ are significantly smaller than
3. At low compactness, however, the crust becomes important because
it constitutes a larger fraction of the NS, since the crust-core
interface density $\rho_0$ is attained deeper inside the star. The
softness of the crust is responsible for the flattening of the Love
number curves observed for $M/R \lesssim 0.15$.

\section{I-Love relations}\label{Sec:Universality}

As we described in the introduction, universal relations between
macroscopic properties of NSs have emerged as a promising tool for
observational astrophysics and gravitational-wave astronomy. To date,
these relations have implicated the NS moment of inertia, the
scaled tidal Love numbers, and the spin-induced quadrupole
moment. Here, we extend the universal I-Love relations to include also
the scaled \emph{rotational-tidal} Love numbers.

\begin{figure}[tb]
\begin{subfigure}[b]{0.45\linewidth}
\includegraphics[]{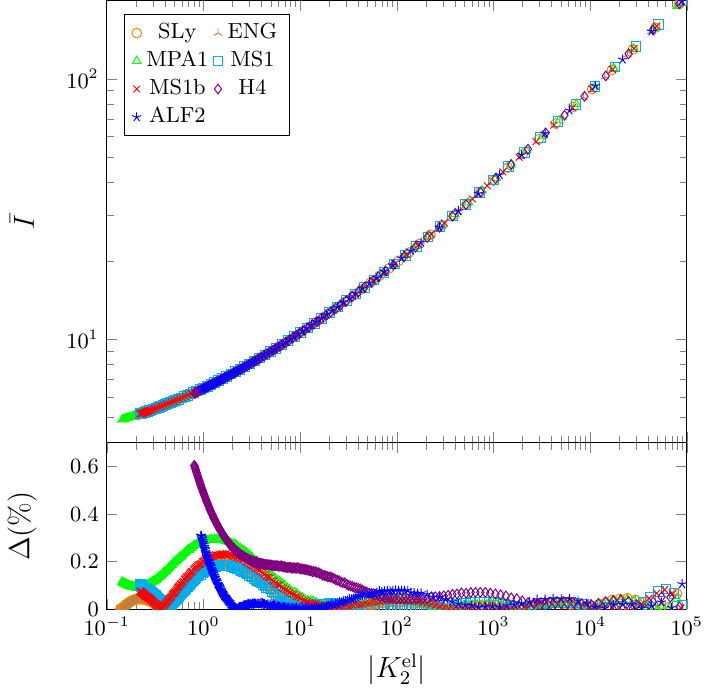} 
\caption{$\bar{I}$--$K_2^\text{el}$ relation}
\label{fig:ILQ-K_el}
\end{subfigure}
\begin{subfigure}[b]{0.45\linewidth}
\includegraphics[]{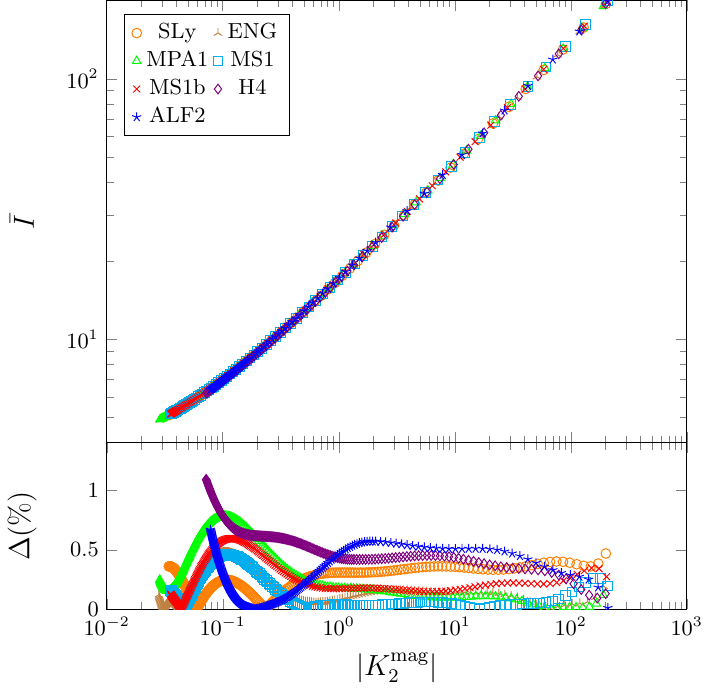} 
\caption{$\bar{I}$--$K_2^\text{mag}$ relation}
\label{fig:ILQ-K_mag}
\end{subfigure}
\begin{subfigure}[b]{0.45\linewidth}
\includegraphics[]{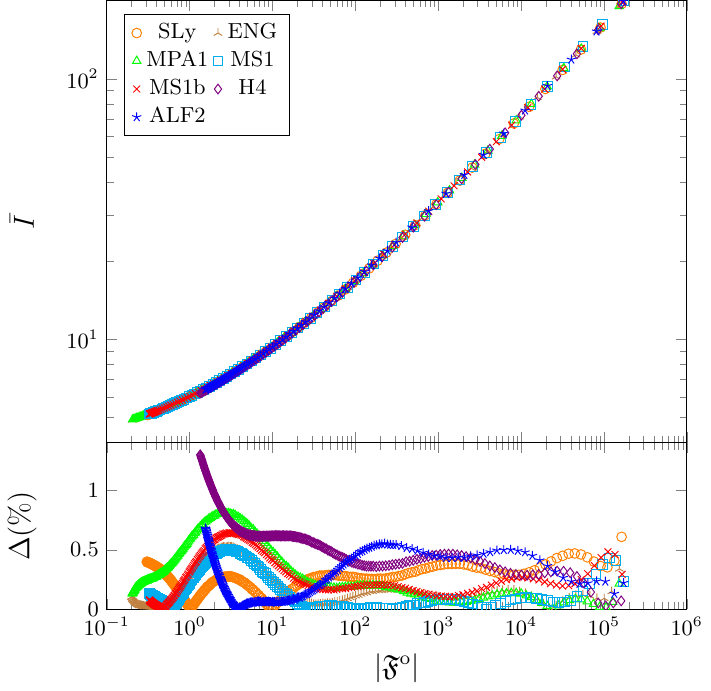} 
\caption{$\bar{I}$--$\mathfrak{F}^\text{o}$ relation}
\label{fig:ILQ-F_o}
\end{subfigure}
\begin{subfigure}[b]{0.45\linewidth}
\includegraphics[]{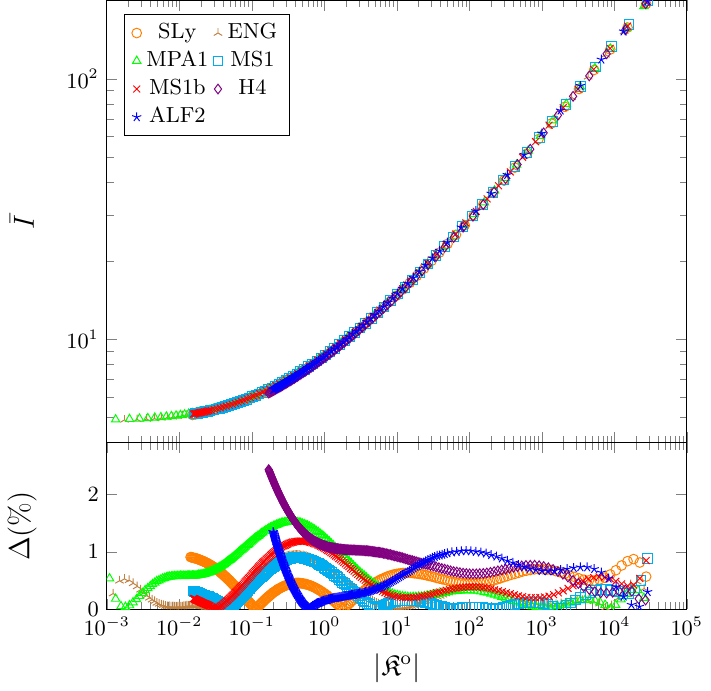} 
\caption{$\bar{I}$--$\mathfrak{K}^\text{o}$ relation}
\label{fig:ILQ-K_o}
\end{subfigure}
\caption{I-Love relations for the scaled Love numbers
  $\{ K_2^\text{el}, K_2^\text{mag}, \mathfrak{F}^{\text{o}},
  \mathfrak{K}^{\text{o}} \}$ calculated with the seven realistic EoS
  models described in Sec.~\ref{Sec:RealEoS}. Note that the astrophysical range of interest
    corresponds to $4<\bar{I}<30$, but we
    include larger values of $\bar{I}$ for comparison with other
    analyses. Insets show the
  deviations $\Delta$ from universality (in $\%$) for each EoS with respect to fits of
  the form of Eq.~(\ref{eq:fit}) with the coefficients of
  Table~\ref{tb:fit}.}
\label{fig:ILQ}
\end{figure}

We seek EoS-independent functional relationships between the
dimensionless moment of inertia $\bar{I} \equiv I/M^3$ and each of the scaled
Love numbers
$\mathfrak{L} \in \{K^{\text{el}}_2, K^{\text{mag}}_2,
\mathfrak{F}^{\text{o}}, \mathfrak{K}^{\text{o}} \}$. By plotting
$\bar{I}$ against $\mathfrak{L}$ for every EoS in our
sample\footnote{There is no established convention in the literature
  regarding whether $\bar{I}$ or $\mathfrak{L}$ should be taken as the
  independent variable; for instance, Ref.~\cite{ILQ} plots $\bar{I}$
  in terms of $K_2^{\text{el}}$, but Refs.~\cite{Delsate,Pani} plot
  the reverse. We choose to adopt the former arrangement here. This
  should be kept in mind as we compare the numerical values of the
  I-Love deviations with the latter references.}, as in
Fig.~\ref{fig:ILQ}, and performing a log-log polynomial fit
\begin{equation}\label{eq:fit}
\log_{10}{\bar{I}_{\text{fit}}} = \sum_{n=0}^{10} c_n \left( \log_{10}{\mathfrak{L}} \right)^n ,
\end{equation}
we can assess a given relation's degree of universality through the
deviations
\begin{equation} \label{eq:dispersion}
  \Delta(\%) = \frac{|\bar{I}-\bar{I}_{\text{fit}}|}{\bar{I}_{\text{fit}}}\times 100
\end{equation}
from the fit. The coefficients $c_n$ of our fits are listed in
Table~\ref{tb:fit}, and the deviations are plotted in the insets of
Fig.~\ref{fig:ILQ}.

We observe deviations from the universal $\bar{I}$--$K_2^{\text{el}}$
relation of less than $0.7 \%$, and of $0.1\%$ on average, for our
sample of EoSs. This is broadly consistent with the results reported
in the literature \cite{ILQ,Pani}, which have maximum deviations of
order $1\%$. For the $\bar{I}$--$K_2^{\text{mag}}$ fit, we find
maximum deviations of $1.1\%$ (average deviations of $0.3\%$);
residuals of less than $5\%$ ($2\%$ on average) were reported in
Ref.~\cite{Delsate} for NSs in the irrotational fluid state. The
universality is weaker for the static fluid state: Refs.~\cite{Pani}
and \cite{Delsate} observed maximum deviations of 6\% and 10\%,
respectively, in this case. We attribute our slightly smaller
deviations overall to differences in the sample of EoSs used.

\begin{table}
  \begin{ruledtabular}
    \centering
    \begin{tabular}{ M{0.15\textwidth} M{0.2\textwidth} M{0.2\textwidth} M{0.2\textwidth} M{0.2\textwidth} }
      Coefficient & $\bar{I}$--$K_2^\text{el}$ & $\bar{I}$--$K_2^\text{mag}$ & $\bar{I}$--$\mathfrak{F}^{\text{o}}$ & $\bar{I}$--$\mathfrak{K}^{\text{o}}$\\ 
      \hline \noalign{\medskip}
      $c_0$  & $\phantom{-}8.115 \times 10^{-1}$ & $\phantom{-}1.232 \phantom{\;\times 10^{-1}}$ & $\phantom{-}7.782 \times 10^{-1}$ & $\phantom{-}9.325 \times 10^{-1}$ \\
      $c_1$ & $\phantom{-}1.763 \times 10^{-1}$ & $\phantom{-}4.242 \times 10^{-1}$ & $\phantom{-}1.518 \times 10^{-1}$ & $\phantom{-}2.035 \times 10^{-1}$\\
      $c_2$ & $\phantom{-}3.995 \times 10^{-2}$ & $\phantom{-}2.635 \times 10^{-2}$ & $\phantom{-}3.849 \times 10^{-2}$ & $\phantom{-}4.147 \times 10^{-2}$ \\
      $c_3$ & $-9.958 \times 10^{-4} $ & $-3.934 \times 10^{-3}$ & $\phantom{-}7.846 \times 10^{-3}$ & $-7.056 \times 10^{-3}$ \\
      $c_4$ & $-3.734 \times 10^{-3}$ & $-1.171 \times 10^{-4}$  & $-4.380 \times 10^{-3}$ & $-1.255 \times 10^{-3}$ \\
      $c_5$ & $\phantom{-}1.005 \times 10^{-3}$ & $-3.914 \times 10^{-3}$  & $-6.483 \times 10^{-3}$ & $\phantom{-}1.292 \times 10^{-3}$ \\
      $c_6$ & $\phantom{-}9.147 \times 10^{-4}$ & $\phantom{-}2.898 \times 10^{-3}$  & $\phantom{-}7.131 \times 10^{-3}$ & $\phantom{-}3.647 \times 10^{-5}$ \\
      $c_7$ & $-7.091 \times 10^{-4}$ & $\phantom{-}3.935 \times 10^{-4}$  & $-3.014 \times 10^{-3}$ & $-1.413 \times 10^{-4}$ \\
      $c_8$ & $\phantom{-}2.018 \times 10^{-4}$ & $-7.399 \times 10^{-4}$  & $\phantom{-}6.544 \times 10^{-4}$ & $\phantom{-}8.501\times 10^{-6} $ \\
      $c_9$ & $-2.674 \times 10^{-5}$ & $\phantom{-}1.219 \times 10^{-4}$ & $-7.247 \times 10^{-5}$ & $\phantom{-}6.347\times 10^{-6} $ \\
      $c_{10}$ & $\phantom{-}1.377\times 10^{-6}$ & $\phantom{-}1.010 \times 10^{-5}$ & $\phantom{-}3.249\times 10^{-6}$ & $-8.400\times 10^{-7}$ \\
    \end{tabular}
  \end{ruledtabular}
  \caption{Coefficients of the $\bar{I}$--$\mathfrak{L}$ fit for each scaled
    Love number.}
  \label{tb:fit}
\end{table}
\begin{table}
  \begin{ruledtabular}
    \centering
    \begin{tabular}{ M{0.15\textwidth} M{0.2\textwidth} M{0.2\textwidth} M{0.2\textwidth} }
      Coefficient & $K_2^\text{mag}$--$K_2^{\text{el}}$ & $\mathfrak{F}^{\text{o}}$--$K_2^{\text{el}}$ & $K_2^{\text{el}}$--$\mathfrak{K}^{\text{o}}$\\ 
      \hline \noalign{\medskip}
      $c_0$  & $-1.098 \phantom{\;\,\times 10^{-1}}$ & $\phantom{-}2.083 \times 10^{-1}$ & $\phantom{-}6.083 \times 10^{-1}$ \\
      $c_1$ & $\phantom{-}5.644 \times 10^{-1}$ & $\phantom{-}1.050 \phantom{\;\,\times 10^{-2}}$ & $\phantom{-}9.359 \times 10^{-1}$ \\
      $c_2$ & $\phantom{-}2.089 \times 10^{-2}$ & $-5.389 \times 10^{-2}$ & $\phantom{-}5.264 \times 10^{-2}$ \\
      $c_3$ & $-1.826 \times 10^{-2}$ & $-2.082 \times 10^{-2}$ & $-8.223 \times 10^{-2}$ \\
      $c_4$ & $\phantom{-}4.358 \times 10^{-2}$ & $\phantom{-}9.608 \times 10^{-2}$ & $\phantom{-}1.131 \times 10^{-2}$ \\
      $c_5$ & $-4.877 \times 10^{-3}$ & $-2.342 \times 10^{-2}$ & $\phantom{-}4.248 \times 10^{-2}$ \\
      $c_6$ & $-3.331 \times 10^{-2}$ & $-7.012 \times 10^{-2}$ & $\phantom{-}1.054 \times 10^{-3}$ \\ 
      $c_7$ & $\phantom{-}2.430 \times 10^{-2}$ & $\phantom{-}6.092 \times 10^{-2}$ & $-1.276 \times 10^{-2}$ \\
      $c_8$ & $-5.740 \times 10^{-3}$ & $-1.850 \times 10^{-2}$ & $-1.818 \times 10^{-3}$ \\
      $c_9$ & $\phantom{-}3.740 \times 10^{-5}$ & $\phantom{-}1.712 \times 10^{-3}$ & $\phantom{-}1.345 \times 10^{-3}$ \\
      $c_{10}$ & $\phantom{-}1.078 \times 10^{-4}$ & $\phantom{-}7.841 \times 10^{-5}$ & $\phantom{-}3.018 \times 10^{-4}$ \\
      \hline
      \hline
      \vspace{2pt} 
      Coefficient & $K_2^\text{mag}$--$\mathfrak{K}^{\text{o}}$& $\mathfrak{F}^{\text{o}}$--$\mathfrak{K}^{\text{o}}$ & $\mathfrak{F}^{\text{o}}$--$K_2^\text{mag}$\\ 
      \hline \noalign{\medskip}
      $c_0$  & $-7.455 \times 10^{-1}$ & $\phantom{-}8.324 \times 10^{-1}$ & $\phantom{-}2.022 \phantom{\;\,\times 10^{-1}}$ \\
      $c_1$ & $\phantom{-}5.566 \times 10^{-1}$ & $\phantom{-}9.474 \times 10^{-1}$ & $\phantom{-}1.511 \phantom{\;\,\times 10^{-1}}$ \\
      $c_2$ & $\phantom{-}5.909 \times 10^{-2}$ & $\phantom{-}4.750 \times 10^{-2}$ & $-9.236 \times 10^{-2}$ \\
      $c_3$ & $-3.377 \times 10^{-2}$ & $-5.985 \times 10^{-2}$ & $-5.106 \times 10^{-3}$ \\
      $c_4$ & $\phantom{-}4.361 \times 10^{-3}$ & $\phantom{-}9.396 \times 10^{-3}$ & $-2.805 \times 10^{-2}$ \\
      $c_5$ & $\phantom{-}1.585 \times 10^{-2}$ & $\phantom{-}2.685 \times 10^{-2}$ & $\phantom{-}1.619 \times 10^{-1}$ \\
      $c_6$ & $\phantom{-}2.551 \times 10^{-4}$ & $\phantom{-}1.023 \times 10^{-3}$ & $\phantom{-}2.206 \times 10^{-1}$ \\ 
      $c_7$ & $-4.713 \times 10^{-3}$ & $-8.216 \times 10^{-3}$ & $-1.670 \times 10^{-1}$ \\
      $c_8$ & $-6.642 \times 10^{-4}$ & $-1.275\times 10^{-3}$ & $-4.769 \times 10^{-1}$ \\
      $c_9$ & $\phantom{-}4.981 \times 10^{-4}$ & $\phantom{-}8.753 \times 10^{-4}$ & $-3.085 \times 10^{-1}$ \\
      $c_{10}$ & $\phantom{-}1.118 \times 10^{-4}$ & $\phantom{-}2.021 \times 10^{-4}$ & $-6.576 \times 10^{-2}$ \\
    \end{tabular}
  \end{ruledtabular}
  \caption{Coefficients for fits of the same form as
    Eq.~(\ref{eq:fit}) between each pair of scaled Love numbers $\mathfrak{L}_1$--$\mathfrak{L}_2$, with $\mathfrak{L}_2$ used as the independent variable. As noted in
    Footnote~\ref{foot:Ko_fit}, $\mathfrak{K}^{\text{o}}$ is used as
    the independent variable when it appears.}
  \label{tb:fit2}
\end{table}

We find that the extended I-Love relations involving the scaled
rotational-tidal Love numbers are also nearly EoS-independent,
although the degree of universality is slightly weaker than for the scaled
tidal Love numbers. For the $\bar{I}$--$\mathfrak{F}^{\text{o}}$ fit,
deviations from universality average to $0.3\%$, with a maximum of
$1.3\%$, whereas for the $\bar{I}$--$\mathfrak{K}^{\text{o}}$ fit they
average to $0.6\%$, and are always smaller than $2.5\%$. These results
are in sharp contrast to Ref.~\cite{Pani}, which reported that I-Love
universality was \emph{broken} by the scaled rotational-tidal Love
numbers---deviations of order 200\% and 50\% were found for
$\delta \tilde{\lambda}_M^{(32)}$ (our $\mathfrak{F}^\text{o}$) and
$\delta \tilde{\lambda}_E^{(32)}$ (our $\mathfrak{K}^\text{o}$),
respectively, for NSs in the static state. We suspect that a
discrepancy of this magnitude is not simply due to the difference in
fluid state, and is likely rooted in the same problem that caused our
disagreement for polytropes.

We observe that for the scaled tidal Love numbers, I-Love universality is weaker in the gravitomagnetic sector, as claimed by Ref.~\cite{Pani}, and we see that the trend persists for the scaled rotational-tidal Love numbers. We also checked that for polytropes, softer EoSs depart more
strongly from the realistic EoS fit (not shown
in Fig.~\ref{fig:ILQ}), extending existing intuition from the case of the scaled
tidal Love numbers~\cite{ILQ}. Indeed, for polytropes with $n=0.5$,
$n=0.75$, and $n=1$, we get deviations from the
$\bar{I}$--$\mathfrak{F}^{\text{o}}$ fit of less
than $1.4\%$, $1.75\%$, and $3.7\%$, respectively; for the
$\bar{I}$--$\mathfrak{K}^{\text{o}}$ fit, the deviations are less than
$2.7\%$, $3.1\%$, and $6.0\%$. We note that the degree of universality
is again stronger in the gravitoelectric sector.

\begin{figure}
  \begin{subfigure}[b]{0.45\linewidth}
    \includegraphics[]{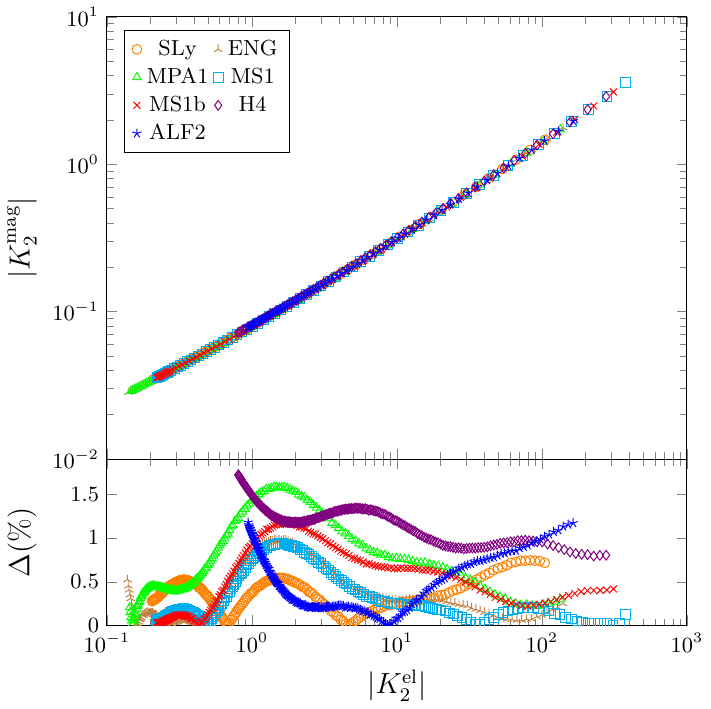} 
    \caption{$K_2^\text{mag}$--$K_2^{\text{el}}$ relation}
    \label{fig:LL-K_mag}
  \end{subfigure}
  \begin{subfigure}[b]{0.45\linewidth}
    \includegraphics[]{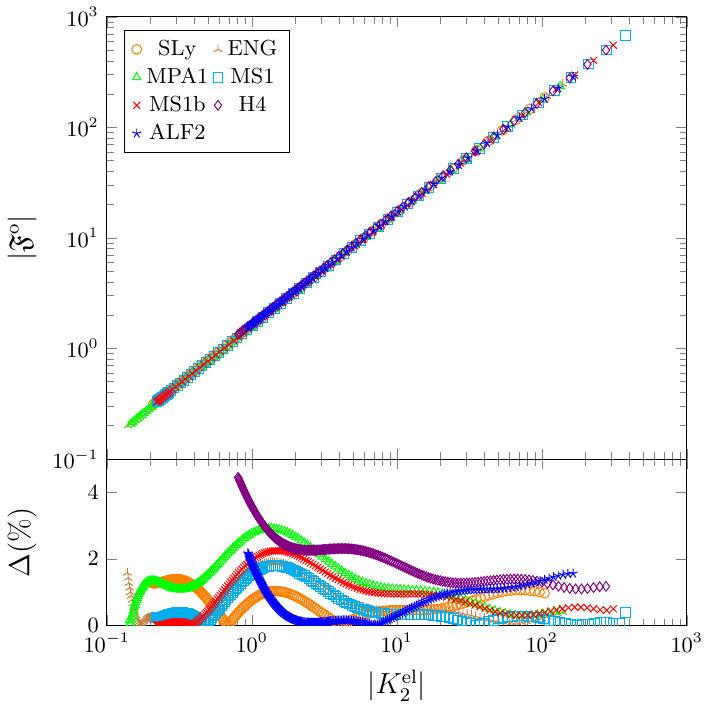} 
    \caption{$\mathfrak{F}^\text{o}$--$K_2^{\text{el}}$ relation}
    \label{fig:LL-F_o}
  \end{subfigure}
  
  \begin{subfigure}[b]{0.45\linewidth}
    \includegraphics[]{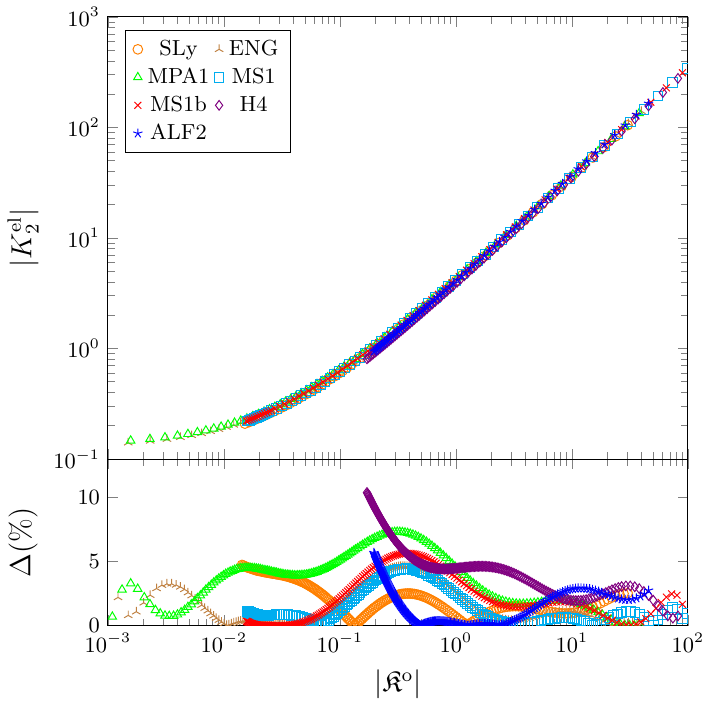} 
    \caption{$K_2^{\text{el}}$--$\mathfrak{K}^\text{o}$ relation}
    \label{fig:LL-K_o}
  \end{subfigure}
    \begin{subfigure}[b]{0.45\linewidth}
    \includegraphics[]{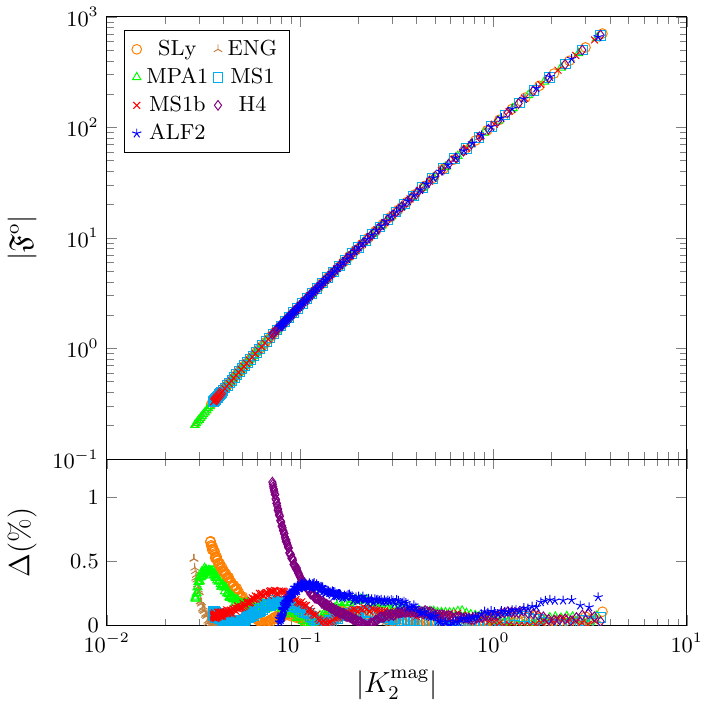} 
    \caption{$\mathfrak{F}^\text{o}$--$K_2^{\text{mag}}$ relation}
    \label{fig:F_o-K_mag}
  \end{subfigure}
  
      \begin{subfigure}[b]{0.45\linewidth}
    \includegraphics[]{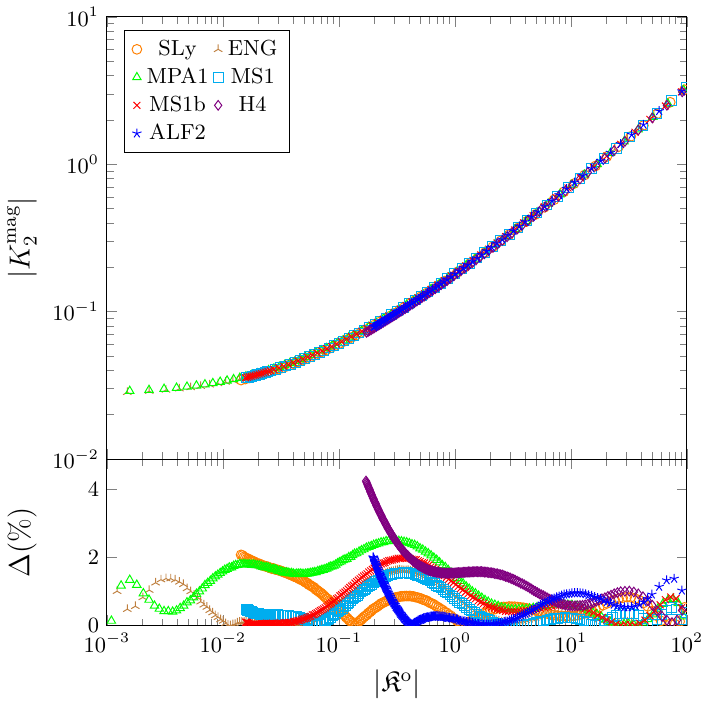} 
    \caption{$K_2^{\text{mag}}$--$\mathfrak{K}^\text{o}$ relation}
    \label{fig:K_mag-K_o}
  \end{subfigure}
    \begin{subfigure}[b]{0.45\linewidth}
    \includegraphics[]{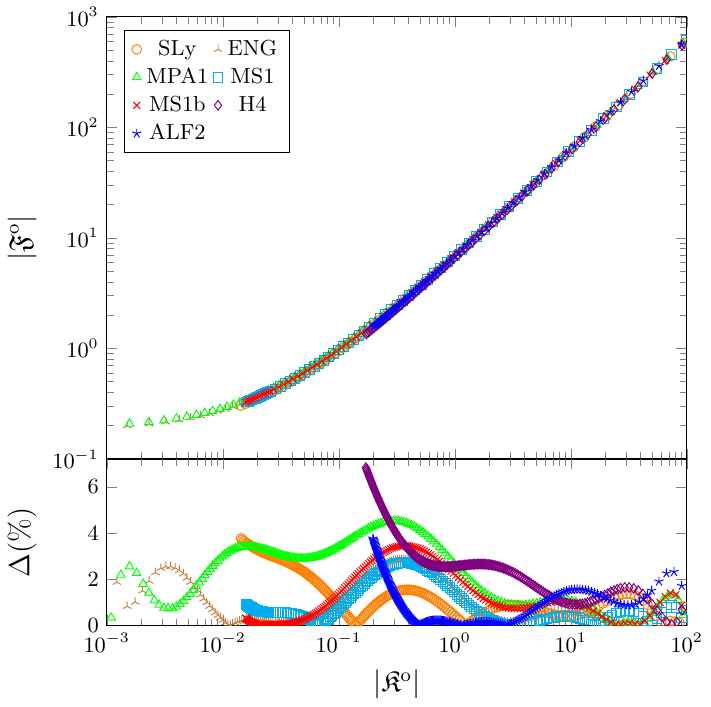} 
    \caption{$\mathfrak{F}^\text{o}$--$\mathfrak{K}^\text{o}$ relation}
    \label{fig:F_o-K_o}
  \end{subfigure}
  
  \caption{Universal relations between each pair of scaled Love numbers. Insets show the
    deviations $\Delta$ from universality (in~\%) for each EoS with
    respect to fits of the form of Eq.~(\ref{eq:fit}) with the
    coefficients of Table~\ref{tb:fit2}.}
  \label{fig:LoveLove}
\end{figure}

As a consequence of the extended I-Love relations, we also expect
\emph{Love-Love} universal relations to hold between each \emph{pair}
of scaled Love numbers, generalizing the universal
$K_2^{\text{mag}}$--$K_2^{\text{el}}$ relation~\cite{ILQmulti}. We
check these relations in
Fig.~\ref{fig:LoveLove}.

In Fig.~\ref{fig:LL-K_mag} we find that
$K_2^{\text{mag}}$--$K_2^{\text{el}}$ universality holds to within
$1.8\%$ ($0.6\%$ on average) for NSs in the irrotational
state. Refs.~\cite{Pani} and \cite{ILQmulti} found corresponding
deviations of less than $3\%$ and $10\%$, respectively, but those
results refer to the static fluid state. Relations between
$K^{\text{el}}_2$ and the scaled rotational-tidal Love numbers (Figs.~\ref{fig:LL-F_o} and \ref{fig:LL-K_o}) are slightly
less universal, with deviations of less than $4.5\%$ ($1\%$ on
average) seen in the $\mathfrak{F}^{\text{o}}$--$K_2^{\text{el}}$ fit,
and maximum deviations of $10.5\%$ ($2.5\%$ on average) for the
$K_2^{\text{el}}$--$\mathfrak{K}^{\text{o}}$ fit.\footnote{The fact
  that $\mathfrak{K}^{\text{o}}$ approaches zero at large values of
  the compactness (see Fig.~\ref{fig:Realistic_EoS}) is problematic
  for a log--log fit. We overcome this issue by taking
  $\mathfrak{K}^{\text{o}}$ to be the independent variable for the fits in which it is involved. \label{foot:Ko_fit}} Similar relations involving
$K^{\text{mag}}_2$ (Figs.~\ref{fig:F_o-K_mag} and~\ref{fig:K_mag-K_o}) show a slightly better
universality, with deviations of less than $1.2\%$ ($0.2\%$ on
average) seen in the $\mathfrak{F}^{\text{o}}$--$K_2^{\text{mag}}$ fit,
and maximum deviations of $4.3\%$ ($0.9\%$ on average) for the
$K_2^{\text{mag}}$--$\mathfrak{K}^{\text{o}}$ fit. Finally, the  $\mathfrak{F}^{\text{o}}$--$\mathfrak{K}^{\text{o}}$ relation (Fig.~\ref{fig:F_o-K_o}) shows maximum deviations of $6.9\%$ ($1.6\%$ on
average). Coefficients for each
of the fits are given in Table~\ref{tb:fit2}.

Given that approximate universality holds between the dimensionless
moment of inertia $\bar{I}$ and each of the scaled Love numbers
$\{K_2^{\text{el}}, K_2^{\text{mag}},
\mathfrak{F}^{\text{o}},\mathfrak{K}^{\text{o}}\}$, we see no reason
to expect it to be broken when other scaled rotational-tidal
Love numbers are included. In particular, it may be worthwhile to
revisit Ref.~\cite{Pani}'s claim that scaled quadrupole
rotational-tidal Love numbers $\mathfrak{F}^{\text{q}}$ and
$\mathfrak{K}^{\text{q}}$ arising from couplings between the NS spin
and an applied \emph{octupolar} tidal field do \emph{not} satisfy
universal I-Love relations. We conjecture that all of
Sec.~\ref{Sec:TidScales}'s corrections to the leading-order
gravitoelectric quadrupole tides are approximately expressible
in terms of a \emph{single} scaled Love number, for example~$K_2^{\text{el}}$.

Finally, we remark that the original I-Love-Q study \cite{ILQ} also
covered relations involving the rotational quadrupole moment $Q$; we
do not investigate these relations here since $Q$ is second order in
spin, and our perturbative calculation is limited to linear
order. Nevertheless, we expect universal Love-Q relations for scaled
rotational-tidal Love numbers to follow from the original I-Q
relations combined with the extended I-Love relations presented here.
 
\section{Discussion}\label{Sec:Conclusions}

In this paper, we studied the Love numbers of slowly rotating NSs
deformed by weak quadrupolar tides. We computed the rotational-tidal
Love numbers $\mathfrak{f}^{\text{o}}$ and
$\mathfrak{k}^{\text{o}}$---for the first time for realistic NSs in
the irrotational fluid state---as a function of the stellar
compactness for seven chosen EoSs. For astrophysically relevant NS
models, they lie in the ranges
$0.03 \lesssim \mathfrak{f}^{\text{o}} \lesssim 0.22$ and
$0.001 \lesssim \mathfrak{k}^{\text{o}} \lesssim 0.035$. To assist in
future estimates of spin-corrections to tidal effects in NS binaries,
we also provided the Love numbers' specific numerical values for
canonical NSs in Table~\ref{tb:LNs}.

We showed that $\mathfrak{F}^{\text{o}}$ and $\mathfrak{K}^{\text{o}}$
satisfy extended I-Love and Love-Love relations that are universal to
within a few percent in almost every case, in contradiction to previous
work~\cite{Pani}. Despite a different choice of fluid state for the
NS, we compared our results with those of Ref.~\cite{Pani} in the
regime of overlap, and found that they do not agree there. This shows
that the discrepancy in our conclusions is not simply a consequence of
the choice of fluid state. As a check on our computations, we compared
our general-relativistic results for polytropes to Love numbers we
computed in post-Newtonian theory, and we showed that they agree in
the weak-field limit. We also established that our polytrope Love
numbers tend to those of an incompressible fluid as the stiffness of
the EoS is increased. This gives us confidence in our conclusions.

Universality relations tell us that certain seemingly unrelated NS
properties are in fact functionally interdependent. In this paper, we
have shown that the set of interrelated quantities involved in
I-Love universality can be extended to include
$\mathfrak{F}^{\text{o}}$ and $\mathfrak{K}^{\text{o}}$. This extended
I-Love universality supports the idea that---at an approximate
level---NSs are described by just one more parameter than black holes,
regardless of the EoS \cite{ThreeHair}. This simplicity in the
description of NSs is thought to be related to the emergence of an
approximate symmetry in compact stars~\cite{kYlSgPnYtA14}, but a
complete theoretical explanation remains elusive. It would be
interesting to study universality relations involving
quadrupole~\cite{Pani} and higher-$\ell$ scaled rotational-tidal Love numbers
to see if universality holds.

The additional reduction in the parameter space of NSs brought about
by extended I-Love universality is especially useful for
gravitational-wave astronomy. When spin-coupled tidal effects are
eventually incorporated into waveforms, no new parameters will be
needed; to within the accuracy of measurements, it will suffice to use
universality to replace the scaled rotational-tidal Love numbers with
$K_2^{\text{el}}$, the most easily measured Love number. Moreover,
should a rotational-tidal Love number's contribution to the phasing of
the waveform be degenerate with another effect---something like the
spin-spin and rotational quadrupole degeneracy \cite{ILQ}, but at
higher post-Newtonian order---an independent measurement of $I$
(from electromagnetic observations, for instance) or
$K_2^{\text{el}}$ (from gravitational wave measurements) could break the degeneracy.

As the rotational-tidal Love numbers have now been computed for a
variety of candidate NS models, the next task is to determine their
effect on the phase of the inspiral waveform.  Based on their scalings
in the metric, the rotational-tidal deformations can be just as large
as the pure gravitomagnetic ones. Although it is unlikely that this
precise hierarchy of sizes will carry over directly to the tidal
phasing, rotational-tidal effects may well become important for
future-generation detectors. Modeling spin corrections to the tidal
phase is thus timely and important work, and the results presented
here will be essential for that goal.

\acknowledgments
We are glad to thank Eric Poisson for his insightful advice about this paper. We thank Luis Lehner, Kent Yagi, Justin Vines, and Peter Zimmerman for helpful discussions, and we acknowledge Paolo Pani for providing data for comparison. J. G. B. would like to thank the Perimeter Institute for full support via the Undergraduate Student Program. This work was supported in part by the Natural Sciences and Engineering Research Council of Canada. Research at the Perimeter Institute is supported by the Government of Canada through Industry Canada and by the Province of Ontario through the Ministry of Research and Innovation.

\appendix

\section{Post-Newtonian Love numbers} \label{SubSec:PN_calc}
In this section, we calculate the Love numbers $ k_2^\text{el}, k_2^\text{mag}, \mathfrak{f}^\text{o}$, and $\mathfrak{k}^\text{o}$ in post-Newtonian theory. The post-Newtonian results are expected to agree with the general-relativistic Love numbers in the zero-compactness limit. General weak-field formulas for the tidal Love numbers have been derived in the literature: in this regime, the gravitoelectric tidal Love number $k_2^{\text{el}}$ reduces to the Newtonian Love number $k_2$ \cite{Binnington}, which can be calculated with the recipe presented in Sec.~2.4 of Ref.~\cite{Gravity}; and the post-Newtonian gravitomagnetic tidal Love number $k_2^{\text{mag}}$ can be computed via the integral
\begin{equation}
\label{eq:PN_k2_mag}
k_2^\text{mag} = -\frac{2 \pi}{15MR^4}\int_0^R \rho r^6 dr .
\end{equation}
derived by Landry and Poisson in Ref.~\cite{Irrotational}.\footnote{In terms of notation, $k_2^\text{mag}[\text{here}] = \tilde{k}_2^{\text{mag}}[\text{LP}]$. We set their parameter $\lambda$ to 1, which represents the irrotational state that we study here. The fluid quantities in Eq.~\eqref{eq:PN_k2_mag} are solutions to the Newtonian equations of structure, Eq.~\eqref{eq:Newt_eq_struct}.}

The case of the rotational-tidal Love numbers $\mathfrak{f}^{\text{o}}$ and $\mathfrak{k}^{\text{o}}$ has received an incomplete treatment in the literature, and we address it more fully in Secs.~\ref{SubSec:PN_f_o} and~\ref{SubSec:PN_k} below. We derive formulas that are valid for arbitrary barotropic EoSs, including the incompressible fluid model. We also find that $k_2$ is automatically determined as a byproduct of the integration of Eq.~\eqref{eq:U_pert_diff} required to compute $\mathfrak{f}^{\text{o}}$. Throughout this appendix, we restore the physical units of Newton's constant $G$ and the speed of light $c$.

\subsection{Post-Newtonian $\mathfrak{f}^{\text{o}}$}\label{SubSec:PN_f_o}
As a nonlinear phenomenon, the rotational-tidal couplings which give rise to $\mathfrak{f}^{\text{o}}$ and $\mathfrak{k}^{\text{o}}$ are absent in Newtonian theory. Accordingly, a post-Newtonian analysis is required to determine the zero-compactness values of these Love numbers. In this section, we focus on the gravitoelectric rotational-tidal Love number $\mathfrak{f}^{\text{o}}$. The only post-Newtonian calculation of $\mathfrak{f}^{\text{o}}$ in the literature was performed by Landry and Poisson in Ref.~\cite{External_metric} for the special case of an incompressible fluid; they found that
\begin{equation}\label{eq:PN_incompress_f_o}
\mathfrak{f}^{\text{o}}[\text{LP}] = -\bigg(\frac{2GM}{c^2R}\bigg)^5 \mathfrak{F}^{\text{o}}[\text{LP}] = \frac{75}{56} \approx 1.33929 \qquad \text{(incompressible fluid)}
\end{equation}
in the weak-field regime---we recall that they employ a different definition of $\mathfrak{f}^{\text{o}}$ than we do [cf.~Eq.~\eqref{eq:def_f_o_new}]. However, Ref.~\cite{External_metric}'s result disagrees with our general-relativistic calculation of $\mathfrak{f}^{\text{o}}[\text{LP}]$, as is clear from Fig.~\ref{fig:f_o_old}. This discrepancy originates from a term that was missed by Landry and Poisson, as we demonstrate below. (One may note that Eq.~(\ref{eq:PN_incompress_f_o}) gives the post-Newtonian value of $\mathfrak{f}^\text{o}[\text{here}]$; indeed, the term missed in Ref.~\cite{External_metric} happens to correspond precisely to the difference between the definitions of $\mathfrak{f}^\text{o}[\text{here}]$ and $\mathfrak{f}^\text{o}[\text{LP}]$.)

We perform a completely general post-Newtonian calculation of $\mathfrak{f}^{\text{o}}$ for an arbitrary barotropic EoS. Our strategy is as follows: first, we take the unperturbed configuration to be a nonrotating, isolated Newtonian star. Next, we introduce rotational and $\ell = 2$ tidal perturbations. We solve the problem at linear order in the perturbations using Newtonian theory, and then calculate the leading-order terms in the post-Newtonian metric describing the spacetime outside the tidally deformed, rotating star.\footnote{The structure of the tidally deformed, rotating star need not be determined beyond Newtonian order precisely because we are only interested in the leading-order relativistic terms in the post-Newtonian metric. Solving for the structure at 1PN would only contribute higher-order relativistic terms. This discussion is complemented by Footnote~\ref{foot:PN_metric}.} Finally, we compare this post-Newtonian metric to the general-relativistic metric of Sec.~\ref{SubSec:Exterior_solution} in the weak-gravity limit, and we solve for $\mathfrak{f}^{\text{o}}$ algebraically. The remainder of this section is dedicated to presenting these manipulations in detail.

The density profile $\rho$ of a non-rotating, isolated Newtonian star is a solution to the equations of structure for a spherically symmetric ball of fluid, \cite{Gravity}
\begin{equation}\label{eq:Newt_eq_struct}
\frac{dp}{dr} = -\rho \frac{Gm}{r^2}\text{,} \qquad \frac{dm}{dr} = 4 \pi r^2 \rho .
\end{equation}
These equations are supplemented by a barotropic EoS $p=p(\rho)$. Once this background configuration is determined, a velocity field
\begin{equation}\label{eq:v_a}
v^a = \epsilon_{abc}\Omega^b x^c 
\end{equation}
describing rigid rotation is imposed on the unperturbed star. We point out that the density perturbation caused by rotation is proportional to $\Omega^2$ because centrifugal effects are of that order. Therefore, to the approximation level of the general-relativistic perturbation theory, which neglects terms of second order in spin, we can ignore any rotational effects in the density perturbation.

The next step is to introduce the tidal field and to compute the density perturbation that distinguishes the tidally deformed, rotating star from the unperturbed configuration. Since we are only interested in the value of gravitoelectric rotational-tidal Love number $\mathfrak{f}^{\text{o}}$, which is associated with $\mathcal{E}_{ab}$, we turn off the tidal source's gravitomagnetic contribution by setting $\mathcal{B}_{ab} = 0$. Because the gravitoelectric part of the tidal field is simply the usual Newtonian gravitational field, the density perturbation is therefore the same as in the Newtonian theory of linear tides---see, for instance, Sec.~2.4 of Ref.~\cite{Gravity}. However, Ref.~\cite{Gravity} only solves the problem outside the star, while here we must solve the internal problem for the tidally induced density perturbation.

A pure tidal quadrupole field in Newtonian gravity has the gravitational potential
\begin{equation}\label{eq:U_tidal}
U^{\text{tid}} = -\frac{1}{2} \mathcal{E}_{ab}x^ax^b,
\end{equation}
where $\mathcal{E}_{ab} \equiv -\partial_{ab}U^\text{ext}(0)$ is a constant STF tensor comprising second derivatives (evaluated at the origin) of the total gravitational potential $U^\text{ext}$ produced by the tidal source. The deformation induced in response to $U^{\text{tid}}$ is measured by the well-known Newtonian Love number $k_2$, and we compute the tidal perturbations $\delta U$ and $\delta \rho$ of the body's gravitational potential and density in terms of this Love number. Because the perturbations are proportional to the applied tidal field, for purely quadrupolar linear tides we write $\delta U = \tilde{U}(r) \mathcal{E}^\text{q}$, where $\mathcal{E}^\text{q} \equiv \mathcal{E}_{ab}n^an^b$ according to Table~\ref{tb:bili_pot}. Poisson's equation then implies that $\tilde{U}$ satisfies the ODE
\begin{equation}\label{eq:U_pert_diff}
r^2 \frac{d^2\tilde{U}}{dr^2} + 2r \frac{d\tilde{U}}{dr} - \left( 6 + \frac{4\pi r^4 \rho'}{m} \right) \tilde{U} = -\frac{2 \pi r^6 \rho'}{m}.
\end{equation}
Outside the star, the solution to this equation is $\tilde{U} = -k_2 R^5/r^3$ \cite{Gravity}. In the interior, Eq.~\eqref{eq:U_pert_diff} must be solved numerically. A local analysis near $r=0$ reveals that $\tilde{U}'(0)=\tilde{U}(0)=0$ is required for regularity. We therefore solve Eq.~\eqref{eq:U_pert_diff} using a shooting method, imposing the regularity conditions and the continuity of $\tilde{U}(r)$ and $\tilde{U}'(r)$ across $r=R$. The solution automatically determines the value of the Newtonian Love number, $k_2=- \tilde{U}(R)/R^2$, as a consequence of the matching conditions at the stellar surface.

The density perturbation is given by Eq.~(2.218) of Ref.~\cite{Gravity}, $\delta \rho = - \xi^j \partial_j \rho = - \xi^r \rho'$, where the second equality follows from the spherical symmetry of the unperturbed configuration. The radial component of the Lagrangian displacement vector $\xi^j$ is determined to be
\begin{align}\label{eq:Lagrangian_displacement}
\xi^r = \frac{r^2}{Gm} \left( \tilde{U} - \frac{r^2}{2} \right) \mathcal{E}^\text{q}
\end{align}
via the perturbed Euler equation (see e.g.~Eq.~(2.214) of Ref.~\cite{Gravity}). The density perturbation is therefore
\begin{equation}\label{eq:perturbed_density}
\delta \rho = -\frac{\rho'r^2}{Gm} \left( \tilde{U} - \frac{r^2}{2} \right) \mathcal{E}^\text{q}.
\end{equation}

With Eqs.~(\ref{eq:v_a}) and (\ref{eq:perturbed_density}), along with the radial density profile $\rho(r)$ obtained from Eq.~(\ref{eq:Newt_eq_struct}), the relevant part of the structure of the tidally deformed, rotating star is known up to order $\chi^a\mathcal{E}_{bc}$. We are now in a position to compute the contribution of the gravitoelectric rotational-tidal effects to the post-Newtonian metric outside the star. Since we are solely interested in the Love number $\mathfrak{f}^{\text{o}}$, which multiplies the bilinear potential $\mathcal{F}_A^\text{o}$ (defined in Table~\ref{tb:bili_pot} modulo a hidden factor of $c^2$) in the general-relativistic metric of Eq.~\eqref{eq:interior_metric}, we only keep track of the contributions proportional to $\mathcal{F}_a^\text{o}$ (the Cartesian version of the potential; $\mathcal{F}_A^\text{o} = \mathcal{F}_a^\text{o} n^a_A$) in our calculation and we ignore all other terms. Moreover, it turns out that the gauge transformation required to properly compare the post-Newtonian and general-relativistic metrics does not affect the terms involved in the calculation of $\mathfrak{f}^\text{o}$.

To leading order, the post-Newtonian metric components are given in quasi-Cartesian coordinates by\footnote{It can be seen here that post-Newtonian corrections to the structure, which are of order $c^{-2}$ according to the post-Newtonian Euler equation---see e.g. Eq.~(8.119) in Ref.~\cite{Gravity}---would add contributions of order $c^{-4}$ and $c^{-5}$ to the metric.\label{foot:PN_metric}}
\begin{equation}\label{eq:full_PN_metric}
g_{tt} =-1+\frac{2}{c^2}U + {\cal O}(c^{-4})\text{,} \qquad g_{ta} = -\frac{4}{c^3}U_a + {\cal O}(c^{-5}), \qquad g_{ab} = \delta_{ab}\bigg(1+\frac{2}{c^2}U\bigg) + {\cal O}(c^{-4}),
\end{equation}
where $U$ is the Newtonian gravitational potential and $U_a$ is a vector potential obeying the field equation $\nabla^2 U_a = -4 \pi G \rho v_a$. The tidally induced perturbations to this metric involve $\delta U_a$, the bilinear perturbation of the vector potential, which satisfies

\begin{equation}
\nabla^2 \delta U_a=-4 \pi G (\delta_1 j_a +\delta_2 j_a)
\end{equation}
with $\delta_1 j_a \equiv \delta\rho \, v_a$ and $\delta_2 j_a \equiv \rho \, \delta v_a$. Here, $\delta v_a$ is the tidal perturbation of the velocity field $v^a$. The solution to this equation is the sum of the contributions sourced by each $\delta_i j_a $, namely
\begin{equation}\label{eq:pert_vector_pot_int}
\delta_i U_a = G \int \frac{\delta_i j_a(\bf{x}')}{|\bf{x}-\bf{x}'|} d^3x'.
\end{equation}
Only $\delta_1 U_a$ gives a contribution proportional to $\mathcal{F}_a^\text{o}$, so $\delta v_a$ is irrelevant for our purposes. Following the techniques described in Ref.~\cite{External_metric}, which are crucial for solving Eq.~(\ref{eq:pert_vector_pot_int}) analytically, we obtain
\begin{equation}\label{eq:delta_U}
\delta_1 U_a = -\frac{4\pi Gc}{7}\frac{B}{r^4} \mathcal{F}_a^\text{o}, \qquad \text{with} \qquad B \equiv -\int_0^\infty \frac{\rho' r^8}{m} \left( \tilde{U} - \frac{r^2}{2} \right) dr,
\end{equation}
where we have left out any terms not proportional $\mathcal{F}_a^\text{o}$. Therefore, to leading order, the perturbation in the metric due to gravitoelectric rotational-tidal couplings is
\begin{equation}\label{eq:delta_g_ta}
\delta g_{ta} = -\frac{4}{c^3}\delta U_a=  \frac{16\pi G}{7c^2} \frac{B}{r^4} \mathcal{F}_a^\text{o}.
\end{equation} 
To compare this perturbation to the one appearing in the general-relativistic metric, we must re-express it in quasi-spherical coordinates; doing so, we obtain
\begin{equation}\label{eq:delta_g_tA}
\delta g_{tA} = \delta g_{ta} x^a_A = \frac{16\pi G}{7c^2} \frac{B}{r^3}\mathcal{F}_A^\text{o},
\end{equation}
where $x_A^a=\partial x^a/\partial\theta^A$. We emphasize that this
equality leaves out terms not proportional to
$\mathcal{F}_A^\text{o}$. Comparing Eq.~\eqref{eq:delta_g_tA} to the
$tA$ component of the general-relativistic external metric,
Eq.~\eqref{eq:interior_metric}, we infer that the radial function
$f^\text{o}_t(r)$ multiplying $\mathcal{F}_A^\text{o}$ is
\begin{equation}\label{eq:PN_radial_funct}
f_t^\text{o}[\text{PN}] = \frac{16\pi G}{7c^2} \frac{B}{r^3}
\end{equation}
in post-Newtonian theory. To determine the Love number $\mathfrak{f}^{\text{o}}$, we compare Eq.~(\ref{eq:PN_radial_funct}) to the general-relativistic expression for $f_t^\text{o}$, given in Table~\ref{tb:external_metric}, in the zero-compactness limit. To leading post-Newtonian order, we find
\begin{equation}\label{eq:GR_radial_funct}
f_t^\text{o}(r)[\text{GR}] = - \frac{4G}{Mc^2} \frac{I R^5}{r^3} \mathfrak{f}^\text{o} ,
\end{equation}
where the moment of inertia $I$ can be calculated via the Newtonian formula
\begin{equation}\label{eq:inertia}
I = \frac{8\pi}{3}\int_0^R \rho r^4 dr.
\end{equation}
Equating expressions~\eqref{eq:PN_radial_funct} and \eqref{eq:GR_radial_funct} we obtain the general expression
\begin{equation}\label{eq:PN_f_o}
\mathfrak{f}^{\text{o}} = \frac{4\pi}{7I} \frac{M}{R^5}\int_0^R \frac{\rho' r^8}{m} \left( \tilde{U} - \frac{r^2}{2} \right) dr
\end{equation}
for the post-Newtonian gravitoelectric rotational-tidal Love number. In practice, Eq.~(\ref{eq:PN_f_o}), along with Eqs.~(\ref{eq:Newt_eq_struct}), (\ref{eq:U_pert_diff}) and (\ref{eq:inertia}), must be integrated numerically for a given barotropic EoS.\footnote{We note that an integration by parts can be performed in Eq.~(\ref{eq:PN_f_o}) to better condition it for numerical integration. Furthermore, this step allows one to treat the incompressible fluid in the same way as the other EoSs by eliminating the factor of $\rho'$, which is singular at $r=R$ for a body of uniform density. In all cases, the boundary terms vanish unequivocally since the boundaries of integration originating from Eq.~(\ref{eq:delta_U}) extend over the whole spatial domain.}

In the special case of an incompressible fluid, the discontinuity of the density $\rho$ at the stellar surface induces a jump in the derivative of the potential $\tilde{U}$ at $r=R$, whose contribution must be taken into account when matching the internal and external versions of $\tilde{U}$. In order to evaluate this jump, we integrate Eq.~(\ref{eq:U_pert_diff}) through the stellar surface as described in Appendix~\ref{SubSec:Incompress}, obtaining
\begin{equation}
\frac{d\tilde{U}}{dr}(R^+) - \frac{d\tilde{U}}{dr}(R^-) = \frac{3}{R}\left[\frac{R^2}{2}-\tilde{U}(R)\right].
\end{equation}
With this correction, the incompressible fluid Love number $\mathfrak{f}^{\text{o}}$ can be calculated in post-Newtonian theory via the recipe presented above.

Finally, we identify a mistake made by the authors of Ref.~\cite{External_metric} in their comparison of the post-Newtonian and general-relativistic versions of the metric: they asserted that their radial function $f_4^\text{o}$ reduces to $2(2GM/c^2 r)^{6} \mathfrak{F}^\text{o}[\text{LP}]$ in the weak-field limit. However, it should really reduce to $(2GM/c^2 r)^{6}[2 \mathfrak{F}^\text{o}[\text{LP}] -(10/3)K_2^\text{el}]$. Including this missing term, their post-Newtonian result in Eq.~(\ref{eq:PN_incompress_f_o}) becomes $\mathfrak{f}^{\text{o}}[\text{LP}] \approx 0.08929$, which agrees with our general-relativistic incompressible fluid results from Fig.~\ref{fig:f_o_old}.

\subsection{Post-Newtonian $\mathfrak{k}^{\text{o}}$}\label{SubSec:PN_k}
A post-Newtonian calculation of gravitomagnetic rotational-tidal deformations was performed by Poisson and Dou\c{c}ot in Ref.~\cite{Doucot}. However, the authors were interested in effects other than the specific value of the gravitomagnetic rotational-tidal Love number $\mathfrak{k}^\text{o}$, and consequently they only computed $\mathfrak{k}^\text{o}$ for the special case of an $n=1$ polytrope. Here, we adapt their calculation to a general barotropic EoS. Our starting point is Eqs.~(6.4), (6.5), (6.11b), and~(7.9) of Ref.~\cite{Doucot}. The octupole perturbation $\delta U^{\ell = 3}$ of the post-Newtonian gravitational potential can be decomposed in terms of the radial unit vector $n^a$ as
\begin{equation}\label{eq:PN_U_octupolar}
\delta U^{\ell=3} = \mathsf{U}^\text{o}(r) \mathcal{K}_{abc}n^a n^b n^c.
\end{equation}
This equation involves the bilinear moment $\mathcal{K}_{abc}$ defined in Table~\ref{tb:bili_mom}, along with the post-Newtonian version of the gravitomagnetic tidal quadrupole moment, $\mathcal{B}_{ab} \equiv 2 \epsilon^{cd}_{\phantom c\phantom d(a} \partial^{\phantom a}_{b)c} U_d^\text{ext}(0)$. Here, $\partial_{bc} U_a^\text{ext}(0)$ denotes partial derivatives of the vector potential of the tidal source evaluated at the center of the star. The radial function $\mathsf{U}^\text{o}(r)$ has to satisfy Eq.~(7.11) of Ref.~\cite{Doucot},
\begin{equation}\label{eq:PN_U_octupolar_diff}
r^2\frac{d^2\mathsf{U}^\text{o}}{dr^2} +2r\frac{d\mathsf{U}^\text{o}}{dr} - \bigg( 12 + \frac{4\pi r^4\rho'}{m} \bigg) \mathsf{U}^\text{o} =  \frac{8\pi}{9c^2} \frac{r^7 \rho'}{m}.
\end{equation}
This equation is solved outside the star by setting the derivative $\rho'$ to zero and picking out the solution decaying with $r$, which represents the body's tidal response. Doing so, we find that $\mathsf{U}^\text{o}(r) = \alpha r^{-4}$ in the exterior, for $\alpha = \text{constant}$. In the interior, Eq.~\eqref{eq:PN_U_octupolar_diff} must be solved numerically. Local analysis at the center reveals that the regular solution has $\mathsf{U}^\text{o}(r) \propto r^3$ for sufficiently small $r$. Equation~(\ref{eq:PN_U_octupolar_diff}) can therefore be solved using a shooting method with boundary conditions that match $\mathsf{U}^\text{o}(r)$ and its first derivative to the external solution at $r=R$. With the matching value $\mathsf{U}^\text{o}(R)$ at hand, Eq.~(\ref{eq:PN_U_octupolar}) can be compared at $r=R$ to the octupole perturbation $\delta U^{\ell = 3}_{\text{eff}}$ inferred from the general-relativistic external metric in the weak-field limit,
\begin{equation}
\delta U_\text{eff}^{\ell=3} = -\frac{2}{c^2} \bigg( \frac{2GM}{c^2} \bigg)^5 \mathfrak{K}^\text{o} \frac{I}{Mr^4} \mathcal{K}_{abc}n^a n^b n^c ;
\end{equation}
cf. Eq.~(A4) in Ref.~\cite{Doucot}. Here, $I$ is the Newtonian moment of inertia. Equating the post-Newtonian and general-relativistic perturbations, we solve algebraically for the post-Newtonian value of the Love number $\mathfrak{k}^\text{o}$ and find
\begin{equation}\label{eq:PN_k_o}
\mathfrak{k}^\text{o} = \frac{Mc^2\mathsf{U}^\text{o}(R)}{2IR}.
\end{equation}

As in the calculation for $\mathfrak{f}^\text{o}$, the discontinuity of the incompressible fluid density distribution at the stellar surface induces a jump in the derivative of the potential $\mathsf{U}^\text{o}$ at $r=R$. The jump is found by integrating Eq.~(\ref{eq:PN_U_octupolar_diff}) through the stellar surface as described in Appendix~\ref{SubSec:Incompress}; this yields
\begin{equation}
\frac{d\mathsf{U}^\text{o}}{dr}(R^+) - \frac{d\mathsf{U}^\text{o}}{dr}(R^-) = -\frac{3}{R} \left[ \mathsf{U}^\text{o}(R) + \frac{2R^3}{9c^2} \right].
\end{equation}
With this correction, the incompressible fluid Love number $\mathfrak{k}^{\text{o}}$ can be calculated in post-Newtonian theory via the recipe presented above.

\section{Incompressible fluid model}\label{SubSec:Incompress}

In this appendix, we adapt the recipe of Sec.~\ref{Sec:LoveNumbers} for computing the Love numbers to the case of an incompressible fluid. Though the model itself is unphysical, we expect its Love numbers to bound those of other barotropic EoSs from above\footnote{The incompressible fluid does not provide the \emph{lowest} upper bound on the Love numbers by any means. Finer upper bounds using the most extreme NS EoS~\cite{cRrR74} compatible with causality have been investigated in Refs.~\cite{Oeveren,hSnY17}.}: one can view the incompressible fluid as the $n \rightarrow 0$ limit of the polytropic models.

In principle, one can employ the method described in Sec.~\ref{SubSec:Interior_solution} to compute the Love numbers associated with \emph{any} barotropic EoS. However, the implementation of this method turns out to require special care for an incompressible fluid. We examine the subtleties related to this case below; our treatment of the gravitoelectric tidal Love number is essentially the same as that of Refs.~\cite{PriceThorne,Campolattaro,Damour_Nagar}, and we take a similar approach in the novel case of the rotational-tidal Love numbers.

The incompressible fluid model has a constant rest mass density $\rho_* > 0$ inside the star, which abruptly drops to zero outside. Hence, its density profile is $\rho(r) = \rho_* \, \Theta(R-r)$, where $\Theta$ denotes the Heaviside step function. In this case, the first law of thermodynamics written as $d\mu = hd\rho$, where $h$ is the fluid's enthalpy, implies that the total energy density $\mu$ is some constant $\mu_*$ in the interior of the star, so it can be expressed as
\begin{equation}\label{eq:mu_incompressible}
\mu = \mu_* \, \Theta(R-r).
\end{equation}
The total mass of the star is then $M = 4\pi \mu_* R^3/3$ by Eq.~(\ref{eq:simplified_EFE1}). Evidently, the radial derivative of the total energy density is singular at the surface of the star. This implies that radial derivatives of other metric and fluid variables will be generically singular at $r=R$. In particular, some of the radial functions which appear in the metric ansatz Eq.~\eqref{eq:interior_metric} and are directly related to the Love numbers have a discontinuous first derivative at the surface.

Let us focus first on the gravitoelectric sector. In order to compute $K_2^{\text{el}}$ and $\mathfrak{F}^{\text{o}}$ using the method described in Sec.~\ref{SubSec:Gravitoelectric}, we have to integrate Eqs.~(\ref{eq:2_5_Phil_paper})-(\ref{eq:4_8_Phil_paper}) in the interior of the star. The presence of the singular derivative $d\mu/dp$ in Eqs.~(\ref{eq:2_9_Phil_paper}) and (\ref{eq:4_8_Phil_paper}) induces derivative discontinuities of the radial functions $e_{tt}^\text{q}$ and $f_t^{\text{o}}$ at the stellar surface, although the functions themselves remain continuous as required by the matching conditions on the spacetime at $r=R$. Thus, corrections due to such discontinuities need to be taken into account in the integration of $e_{tt}^\text{q}$ and $f_t^{\text{o}}$. In order to evaluate these discontinuities, we integrate Eqs.~(\ref{eq:2_9_Phil_paper}) and (\ref{eq:4_8_Phil_paper}) from $R^- \equiv R - \varepsilon$ to $R^+ \equiv R + \varepsilon$, where $\varepsilon > 0$, and we take the limit $\varepsilon \to 0$. In both cases, we get non-vanishing integrals involving terms of the form $\zeta(r)(\mu + p)(d\mu/dp)$ for some radial function $\zeta(r)$. Differentiating Eq.~(\ref{eq:mu_incompressible}) and using the TOV equation, Eq.~(\ref{eq:TOV}), we find
\begin{equation}\label{eq:dmu_dp}
\int_{R^-}^{R^+} \zeta(r)(\mu + p) \frac{d\mu}{dp} dr = \int_{R^-}^{R^+} \zeta(r) r^2 f \frac{ \mu_* \, \delta(r-R) }{m + 4\pi r^3 p} dr = \zeta(R) R^2 f(R)\mu_*/M.
\end{equation}
With the latter result at hand, we obtain the derivative jumps
\begin{eqnarray}
\frac{d e_{tt}^\text{q}}{dr}(R^+) - \frac{d e_{tt}^\text{q}}{dr}(R^-) &=& - \frac{3}{R}e_{tt}^\text{q}(R), \label{eq:2_9_incompress} \\
\frac{d f_t^{\text{o}}}{dr}(R^+) - \frac{d f_t^{\text{o}}}{dr}(R^-) &=& - 3R ~ e^{-2 \psi(R)}  \left[\omega(R) +  1\right] e_{tt}^\text{q}(R) \label{eq:4_8_incompress}
\end{eqnarray}
from Eqs.~(\ref{eq:2_9_Phil_paper}) and (\ref{eq:4_8_Phil_paper}).
For completeness, we note that the absence of the singular term $d\mu/dp$ in Eq.~(\ref{eq:2_5_Phil_paper}) prevents any discontinuity in $d\omega/dr$, and thus no jump correction needs to be incorporated in the integration of that equation.

Next, in order to compute the scaled gravitomagnetic Love numbers $K_2^{\text{mag}}$ and $\mathfrak{K}^{\text{o}}$ following the method described in Sec.~\ref{SubSec:Gravitomagnetic}, we need to integrate Eqs.~(\ref{eq:5_4_Phil_paper})-(\ref{eq:5_16_Phil_paper}) in the interior of the star. As in the case of Eq.~(\ref{eq:2_5_Phil_paper}), no correction is needed in the integration of Eq.~(\ref{eq:5_4_Phil_paper}). However, in principle, one has to account for derivative jumps in the functions $k_{tr1}^{\text{o}}$ and $k_{tt}^{\text{o}}$. We evaluate these jumps by integrating Eqs.~(\ref{eq:5_15_Phil_paper}) and (\ref{eq:5_16_Phil_paper}) from $R^-$ to $R^+$, taking the limit $\varepsilon \to 0$. The new singular terms are either of the form
\begin{equation}
\zeta(r) \frac{d\mu}{dp} k_{tr1}^{\text{o}} \qquad \mbox{or} \qquad \zeta(r) \frac{d\mu}{dp} \frac{dk_{tr1}^{\text{o}}}{dr}.
\end{equation}
In order to evaluate their integrals across the stellar surface, we recall the fact that $k_{tr1}^{\text{o}} = 0$ for $r \geq R$, so we can always replace $k_{tr1}^{\text{o}}$ by $k_{tr1}^{\text{o}} \Theta(R-r)$ independently of its explicit functional form in the interior. The same argument applies to the derivative $dk_{tr1}^{\text{o}}/dr$ and the pressure $p$, so that we can also write $dk_{tr1}^{\text{o}}/dr = (dk_{tr1}^{\text{o}}/dr) \Theta(R-r)$ and $\mu + p = (\mu_* + p)\Theta(R-r)$. This technical artifice leads to cancellations of Heaviside functions in the singular terms. Thus, differentiation of Eq.~(\ref{eq:mu_incompressible}) and substitution of $dp/dr$ from Eq.~(\ref{eq:TOV}) lead to
\begin{equation}
\int_{R^-}^{R^+} \zeta(r) \frac{d\mu}{dp} k_{tr1}^{\text{o}} dr = \int_{R^-}^{R^+} \zeta(r) r^2 f \frac{ \mu_* \, \delta(r-R)}{(\mu_* + p) \left( m + 4\pi r^3 p \right)} k_{tr1}^{\text{o}} dr = \zeta(R) R^2 f(R) k_{tr1}^{\text{o}}(R)/M,
\end{equation}
and likewise for the integral of the singular term involving the derivative $dk_{tr1}^{\text{o}}/dr$. By virtue of the continuity of the function $k_{tr1}^{\text{o}}$ in $r=R$, and the fact that $k_{tr1}^{\text{o}}(R)=0$, from Eq.~(\ref{eq:5_15_Phil_paper}) we obtain the derivative jump
\begin{equation}\label{eq:5_15_incompress2}
\frac{d k_{tr1}^{\text{o}}}{dr}(R^+) - \frac{d k_{tr1}^{\text{o}}}{dr}(R^-) = -  \int_{R^-}^{R^+} \delta(r-R) \frac{d k_{tr1}^{\text{o}}}{dr} dr.
\end{equation}
From the exterior solution we know that $dk_{tr1}^{\text{o}}/dr = 0$ at $r=R$; thus, continuity of $dk_{tr1}^{\text{o}}/dr$ across the surface is sufficient to satisfy Eq.~(\ref{eq:5_15_incompress2}). We therefore conclude that no jump corrections are needed in the integration of Eq.~(\ref{eq:5_15_Phil_paper}). Similarly, when Eq.~(\ref{eq:5_16_Phil_paper}) is integrated across the surface, we get the derivative discontinuity
\begin{equation}\label{eq:5_16_incompress2}
\frac{d k_{tt}^{\text{o}}}{dr}(R^+) - \frac{d k_{tt}^{\text{o}}}{dr}(R^-) = \frac{1}{R} \left[ 2 \left( 2\omega(R) - 3 \right) b_t^{\text{q}}(R) - 3 k_{tt}^{\text{o}}(R) \right].
\end{equation}
With the addition of these jump corrections, the set of Love numbers $\{ K_2^{\text{el}}, K_2^{\text{mag}}, \mathfrak{F}^{\text{o}}, \mathfrak{K}^{\text{o}} \}$ can be computed for an incompressible fluid by following the method described in Sec.~\ref{SubSec:Interior_solution} .

\section{Quadrupole rotational-tidal Love numbers}\label{SubSec:QuadLNs}

Two kinds of scaled quadrupole rotational-tidal Love numbers are alluded to in this paper: the $\ell = 2$ Love numbers that arise from couplings between the NS spin and the tidal \emph{quadrupole} moments $\mathcal{E}_{ab}, \mathcal{B}_{ab}$, which we may call $\mathfrak{E}^{\text{q}}$ and $\mathfrak{B}^{\text{q}}$ following prior work \cite{External_metric,Dynamical_response,Internal_metric}; and the $\ell = 2$ Love numbers that are associated with couplings between the NS spin and the tidal \emph{octupole} moments $\mathcal{E}_{abc}, \mathcal{B}_{abc}$, which we referred to as $\mathfrak{F}^{\text{q}}$ and $\mathfrak{K}^{\text{q}}$ in Sec.~\ref{Sec:Universality}. The latter were studied by Pani, Gualtieri and Ferrari in Ref.~\cite{Pani}, but were omitted in this work. The former were found by Landry \cite{Internal_metric} to possess a universal value of $1/120$ for material bodies, independently of the EoS.

In this appendix, we clarify the nature of the scaled quadrupole rotational-tidal Love numbers $\mathfrak{E}^{\text{q}}$, $\mathfrak{B}^{\text{q}}$ generated by the external tidal quadrupole. The scaled Love number $\mathfrak{E}^{\text{q}}$ is associated with the bilinear quadrupole moment $\hat{\mathcal{E}}_{ab} \equiv 2 \chi^c \epsilon_{cd(a}\mathcal{E}^{d}_{\phantom{d}b)}$ introduced in Ref.~\cite{External_metric}. This bilinear moment makes its appearance in the metric ansatz of Eq.~\eqref{eq:interior_metric} through the scalar potential $\hat{\mathcal{E}}^{\text{q}} \equiv \hat{\mathcal{E}}_{ab} n^a n^b$ defined in Table~\ref{tb:bili_pot}. According to Ref.~\cite{External_metric}, an equivalent expression for the potential is
\begin{equation} \label{ehatqalt}
\hat{\mathcal{E}}^{\text{q}} = -\chi \partial_{\phi} \mathcal{E}^{\text{q}} .
\end{equation}

Let us consider the effect on Eq.~\eqref{ehatqalt} of a shift
\begin{equation} \label{shift}
\phi \to \phi - \kappa \chi
\end{equation}
in the angular coordinate $\phi$, where $\kappa$ is an arbitrary parameter. To first order in $\chi$, this corresponds to a change
\begin{equation}
\mathcal{E}^{\text{q}} \to \mathcal{E}^{\text{q}}  + \kappa \hat{\mathcal{E}}^{\text{q}}
\end{equation}
in the tidal potential. The effect of this shift on the external metric ansatz, Eq.~(4.4) of Ref.~\cite{External_metric}, is to take
\begin{align}
\hat{e}^{\text{q}}_{tt} &\to \hat{e}^{\text{q}}_{tt} - \kappa e^{\text{q}}_{tt} , \\
\hat{e}^{\text{q}}_{rr} &\to \hat{e}^{\text{q}}_{rr} - \kappa e^{\text{q}}_{rr} , \\
\hat{e}^{\text{q}} &\to \hat{e}^{\text{q}} - \kappa e^{\text{q}} .
\end{align}
As can be seen from the expressions for the radial functions given in Table IV of Ref.~\cite{External_metric}, this amounts to a shift
\begin{equation}
\mathfrak{E}^{\text{q}} \to \mathfrak{E}^{\text{q}} - \kappa K_2^{\text{el}}
\end{equation}
in the Love number. Hence, $\mathfrak{E}^{\text{q}}$ is gauge-dependent, and can be adjusted to make $\hat{e}^{\text{q}}_{tt}, \hat{e}^{\text{q}}_{rr}$ and $\hat{e}^{\text{q}}$ vanish. In the Regge-Wheeler gauge, this is achieved by setting $\mathfrak{E}^{\text{q}} = 1/120$, as dictated by the interior solution \cite{Internal_metric}.

The shift \eqref{shift} also produces a change in $ \hat{\mathcal{B}}^{\text{q}}_A \equiv \epsilon_{abc} n^b \hat{\mathcal{B}}^c_{\;d} n^d n_A^a = -\chi \partial_{\phi} \mathcal{B}^{\text{q}}_A , $ where the bilinear tidal moment is $\hat{\mathcal{B}}_{ab} \equiv 2 \chi^c \epsilon_{cd(a}\mathcal{B}^{d}_{\phantom{d}b)}$. The net effect is to shift the scaled Love number
\begin{equation}
\mathfrak{B}^{\text{q}} \to \mathfrak{B}^{\text{q}} - \kappa K_2^{\text{mag}} .
\end{equation}
Once $\kappa$ has been selected to set $\mathfrak{E}^{\text{q}}$ to a desired value, the gauge freedom in $\phi$ is exhausted, and $\mathfrak{B}^{\text{q}}$ acquires a physical meaning: Landry has shown that it reflects the presence of an $r$-mode in the star.

Since we have demonstrated that the specific values of $\mathfrak{E}^{\text{q}}$ and $\mathfrak{B}^{\text{q}}$ are gauge-dependent, we conclude that they are not true scaled Love numbers. Rather, $\mathfrak{E}^{\text{q}}$ is a gauge constant associated with the freedom to shift the angular coordinate $\phi$; fixing this freedom sets the value of $\mathfrak{B}^{\text{q}}$ up to a residual physical dependence on internal $r$-modes. The point is that the couplings between $\chi^a$ and $\mathcal{E}_{ab}$, $\mathcal{B}_{ab}$ produce only two actual scaled rotational-tidal Love numbers, $\mathfrak{F}^{\text{o}}$ and $\mathfrak{K}^{\text{o}}$, not four as was claimed by Refs.~\cite{External_metric,Dynamical_response,Internal_metric}. Finally, although this argument is formulated in the Regge-Wheeler gauge, we remark that a similar argument holds in the light-cone gauge \cite{Preston} employed in Sec.~III of Ref.~\cite{External_metric}.

\section{Comparison with Pani, Gualtieri \& Ferrari}\label{SubSec:PGF}

In this appendix, we derive the mappings between our scaled Love numbers $K_2^{\text{el}}$, $K_2^{\text{mag}}$, $\mathfrak{F}^{\text{o}}$, $\mathfrak{K}^{\text{o}}$ and their equivalents in Ref.~\cite{Pani}. A subset of these relations appeared without a detailed derivation in Ref.~\cite{Internal_metric}. They permit us to compare our quantitative results with those of Ref.~\cite{Pani}, as we do in Sec.~\ref{Sec:results}.

Our Love numbers are defined as integration constants associated with decaying solutions in the exterior metric \cite{External_metric}. The Love numbers of Ref.~\cite{Pani} are defined in terms of derivatives of induced multipole moments with respect to tidal moments. Nonetheless, the expressions obtained by Pani, Gualtieri \& Ferrari ultimately involve metric components with integration constants of their own: their scaled tidal Love numbers are

\begin{equation} \label{eq:panitid}
\tilde{\lambda}^{(2)}_E = \frac{2\gamma_2}{\sqrt{5\pi}\alpha_2} , \qquad \tilde{\lambda}_M^{(2)} = \frac{\gamma_2^*}{480 \alpha_2^*}
\end{equation}
and their scaled rotational-tidal Love numbers are

\begin{equation} \label{eq:panitidrot}
\delta \tilde{\lambda}^{(32)}_M = \frac{44 \sqrt{35}\gamma_2+21\gamma_{32}}{28 \sqrt{7\pi}\alpha_2}, \qquad \delta \tilde{\lambda}^{(32)}_E = -\gamma_{32}^*/\alpha_2^* .
\end{equation}
The constants $\alpha_2$, $\alpha_2^*$, $\gamma_2$, $\gamma_2^*$, $\gamma_{32}$, $\gamma_{32}^*$ can be read off the perturbations~(22)-(27) of Ref.~\cite{Pani}. Rather than relating our Love numbers explicitly to multipole moments, we choose to compare coordinate expressions for the metric and directly identify the constants appearing in Eqs.~\eqref{eq:panitid}-\eqref{eq:panitidrot}. In general, this would require a transformation of the metric, but fortunately both works employ Boyer-Lindquist $(t,r,\theta,\phi)$ coordinates and the Regge-Wheeler gauge.

The exterior solution of Ref.~\cite{Pani} is obtained by supplementing its background metric (9) with the perturbations (22)-(27) of that paper.\footnote{Eqs.~(22)-(27) of Ref.~\cite{Pani} are the components relevant for this discussion, but the full metric perturbation can be found in Appendix A of that paper. In practice, we make use of the perturbed metric provided as a \textsc{Mathematica}\,\textregistered ~notebook in the Supplemental Material to Ref.~\cite{Pani} at \url{http://link.aps.org/supplemental/10.1103/PhysRevD.92.124003}.} Because Pani, Gualtieri \& Ferrari restrict themselves to axisymmetric perturbations, we specialize our ansatz \eqref{eq:interior_metric} to axisymmetry by dropping the terms with hatted potentials. We then compare the metric components order-by-order in a perturbative expansion, beginning with the background (which, for the purposes of this appendix, we take to include the rotation). A trivial identification of the coordinates, and the relations $e^{\nu} \to f$, $\mathcal{M} \to M$, $\omega \to \Omega(1-\omega)$, bring their background metric into the same form as ours.

The gravitoelectric tidal perturbations appear in the diagonal components of the metric. In Ref.~\cite{Pani}, the $tt$ component of the perturbation is $\delta g_{tt}^{\text{tid}} = e^{\nu} H_0^{(2)} Y^{20}$, where $Y^{\ell m}(\theta,\phi)$ denotes a spherical harmonic and

\begin{equation}
H_0^{(2)} = \alpha_2 y(y-2) +\gamma_2 \left[ -3 + \frac{1}{2-y} - \frac{1}{y} + 3y + \frac{3}{2}(y-2)y \ln{(1-2y)} \right]
\end{equation}
with $y\equiv r/M$.\footnote{There appears to be a misprint in the sign of the second term in Eq.~(22) in the published version of Ref.~\cite{Pani}. The sign given here matches that found in the Supplemental Material.} Our corresponding metric perturbation is $\delta g_{tt}^{\text{tid}} = e^{\text{q}}_{tt} \mathcal{E}^{\text{q}}$. The exterior solution for $e^{\text{q}}_{tt}$ is given in Table~\ref{tb:external_metric}, and we may expand the tidal potential in spherical harmonics as $\mathcal{E}^{\text{q}} = \sum_m \mathcal{E}^{\text{q}}_m Y^{\ell m}$ \cite{External_metric}. Setting the two perturbations equal, we find that

\begin{equation} \label{eq:params}
\alpha_2 = 4 \sqrt{\frac{\pi}{5}} M^2 \mathcal{E}^{\text{q}}_0 , \qquad \gamma_2 = -32 \sqrt{5\pi} K_2^{\text{el}} M^2 \mathcal{E}^{\text{q}}_0 .
\end{equation}
These relations imply the mapping

\begin{equation}
\tilde{\lambda}^{(2)}_E = -16 \sqrt{\frac{5}{\pi}} K_2^{\text{el}}
\end{equation}
for the scaled gravitoelectric Love number. One can verify that with these associations the $rr$ and $AB$ components of the gravitoelectric tidal perturbation also match exactly.

The gravitoelectric rotational-tidal perturbation appears only in the $tA$ components of the metric. The octupole deformation is given as $\delta g_{tA}^{\ell = 3} = \delta h_0^{(3)} X_A^{30}$ in Ref.~\cite{Pani}; $X_A^{\ell m}=(-\partial_{\phi}Y^{\ell m}/\sin{\theta}, \sin{\theta} \, \partial_{\theta}Y^{\ell m})$ is an odd-parity vector spherical harmonic and

\begin{align}
\delta h_0^{(3)} = &-\frac{M \chi}{6720 y^2} \bigg\{ -128 \sqrt{35}y(5y-4)\alpha_2 \\ \nonumber &- 3 \bigg[8 \sqrt{35} \left(16+44y-90y^2-270y^3+945y^4-405y^5 - y (64-80y+1080y^3-1350y^4+405y^5)\tanh^{-1}{\frac{1}{1-y}} \right) \gamma_2 \\ \nonumber &+ 35y \bigg(8+20y+60y^20-210y^3+90y^4+15y^3(8-10y^2+3y^2)\ln{(1-2/y)}\bigg)\gamma_{32} \bigg] \bigg\} .
\end{align}
Since $\alpha_2$ and $\gamma_2$ are known from Eq.~\eqref{eq:params}, $\gamma_{32}$ is the only undetermined parameter. Our equivalent metric perturbation is $\delta g_{tA}^{\ell = 3} = f^{\text{o}}_t \mathcal{F}^{\text{o}}_A $, with $f^{\text{o}}_t$ given in Table~\ref{tb:external_metric} and $ \mathcal{F}^{\text{o}}_A = \frac{1}{5}\chi \mathcal{E}^{\text{q}}_0 X^{30}_A $ in axisymmetry \cite{External_metric}. The comparison yields

\begin{equation}
\gamma_{32} = \frac{128}{5}\sqrt{\frac{\pi}{7}}\left(\frac{275}{3} K_2^{\text{el}}+28\mathfrak{F}^{\text{o}}\right) M^2 \mathcal{E}^{\text{q}}_0 ,
\end{equation}
and with this association the $\ell = 3 $ gravitoelectric rotational-tidal perturbations agree. It then follows that the scaled gravitoelectric rotational-tidal Love numbers are related by the first of the mappings in Eq.~\eqref{eq:map}. This relation matches the one introduced in Ref.~\cite{Internal_metric}, after taking into account our redefinition of $\mathfrak{F}^{\text{o}}$.

We proceed in a similar fashion in the gravitomagnetic sector. The tidal perturbation appears exclusively in the $tA$ component of the metric, and equating our expression $\delta g_{tA}^{\text{tid}} = b^{\text{q}}_{t} \mathcal{B}^{\text{q}}_A = b^{\text{q}}_{t} \mathcal{B}^{\text{q}}_0 X_A^{20}$ with that of Ref.~\cite{Pani}, $\delta g_{tA}^{\text{tid}} = h_0^{(2)} X_A^{20}$, we identify

\begin{equation}
\alpha_2^* = -\frac{4}{3} \sqrt{\frac{\pi}{5}} M^2\mathcal{B}^{\text{q}}_0 , \qquad \gamma_2^* = -128\sqrt{5\pi} K_2^{\text{mag}}  M^2 \mathcal{B}^{\text{q}}_0 .
\end{equation}
These relations imply that the scaled gravitomagnetic Love number $\tilde{\lambda}_M^{(2)}$ of Ref.~\cite{Pani} is identical to our $K_2^{\text{mag}}$. The rotational-tidal perturbations appear in the $tt$, $rr$ and $AB$ components of the metric. One can show that they all agree  when the assignment

\begin{equation}
\gamma_{32}^* = -\frac{192}{5} \sqrt{7\pi}\mathfrak{K}^{\text{o}} M^2 \mathcal{B}^{\text{q}}_0
\end{equation}
is made, based on the equivalence of our $\delta g_{tt}^{\ell = 3} = k^{\text{o}}_{tt} \mathcal{K}^{\text{o}} = k^{\text{o}}_{tt} \frac{3}{5} \chi \mathcal{B}^{\text{q}}_0 Y^{30}$ and $\delta g_{tt}^{\ell = 3} = e^{\nu}\delta H_0^{(3)} Y^{30}$ of Ref.~\cite{Pani}. This association implies that the second mapping given in Eq.~\eqref{eq:map} relates the scaled gravitomagnetic rotational-tidal Love numbers.

By similar identifications, one can also match the dipole rotational-tidal perturbations, which contain gauge constants rather than Love numbers. Applying the relations derived above, it is straightforward to verify that the exterior metric of Ref.~\cite{Pani} is completely identical to Eq.~\eqref{eq:interior_metric} in axisymmetry. Ultimately, our Love number definitions differ only from those of Pani, Gualtieri and Ferrari by the purely conventional multiplicative factors which we have worked out here.

\bibliography{references}

\end{document}